\documentclass[hyper]{andp2012}

\usepackage[english]{babel}
\usepackage{bbm,color}
\usepackage[alwaysadjust]{paralist}

\makeatletter
\providecommand\@enum@widestlabel{7}
\makeatother

\newcommand{\eqr}[1]{Eq.~(\ref{#1})}
\newcommand{\figr}[1]{Fig.~\ref{#1}}
\newcommand{\secr}[1]{Sec.~(\ref{#1})}
\newcommand{\ii}{{\rm i}}

\definecolor{stephen}{rgb}{1.0,0.1,0.1}

\definecolor{juanjose}{rgb}{0.1,1.0,0.1}

\definecolor{martin}{rgb}{0.1,0.1,1.0}

\definecolor{dieter}{rgb}{0.5,0.5,0.1}

\title{Ultra-fast control of magnetic relaxation in a periodically driven Hubbard model}
\author[J.\,J. Mendoza-Arenas]{Juan Jose Mendoza-Arenas\inst{1,2}\footnote{Corresponding author\quad E-mail:~\textsf{jj.mendoza@uniandes.edu.co}}}
\author[F.\,J. G\'omez-Ruiz]{Fernando Javier G\'omez-Ruiz\inst{1}}
\author[M. Eckstein]{Martin Eckstein\inst{3}}
\author[D. Jaksch]{Dieter Jaksch\inst{2}}
\author[S.\,R. Clark]{Stephen R. Clark\inst{4,3}}

\address[1]{Departamento de F\'isica, Universidad de los Andes, A.A. 4976, Bogot\'a D. C., Colombia}
\address[2]{Clarendon Laboratory, University of Oxford, Parks Road, Oxford OX1 3PU, United Kingdom}
\address[3]{Max Planck Institute for the Structure and Dynamics of Matter, University of Hamburg CFEL, Hamburg, Germany}
\address[4]{Department of Physics, University of Bath, Claverton Down, Bath BA2 7AY, United Kingdom}

\shortauthors{Mendoza-Arenas et al.}
\begin{abstract}
Motivated by cold atom and ultra-fast pump-probe experiments we study the melting of long-range antiferromagnetic order of a perfect N\'eel state in a periodically driven repulsive Hubbard model. The dynamics is calculated for a Bethe lattice in infinite dimensions with non-equilibrium dynamical mean-field theory. In the absence of driving melting proceeds differently depending on the quench of the interactions to hopping ratio $U/\nu_0$ from the atomic limit. For $U \gg \nu_0$ decay occurs due to mobile charge-excitations transferring energy to the spin sector, while for $\nu_0 \gtrsim U$ it is governed by the dynamics of residual quasi-particles. Here we explore the rich effects that strong periodic driving has on this relaxation process spanning three frequency $\omega$ regimes: (i) high-frequency $\omega \gg U,\nu_0$, (ii) resonant $l\omega = U > \nu_0$ with integer $l$, and (iii) in-gap $U > \omega > \nu_0$ away from resonance. In case (i) we can quickly switch the decay from quasi-particle to charge-excitation mechanism through the suppression of $\nu_0$. For (ii) the interaction can be engineered, even allowing an effective $U=0$ regime to be reached, giving the reverse switch from a charge-excitation to quasi-particle decay mechanism. For (iii) the exchange interaction can be controlled with little effect on the decay. By combining these regimes we show how periodic driving could be a potential pathway for controlling magnetism in antiferromagnetic materials. Finally, our numerical results demonstrate the accuracy and applicability of matrix product state techniques to the Hamiltonian DMFT impurity problem subjected to strong periodic driving.
\end{abstract}

\begin{document}
\maketitle

\section{Introduction}
Relaxation of a symmetry-broken state after a quench represents a class of non-equilibrium dynamics that has been intensely studied both experimentally and theoretically. Part of the reason for this interest is the distinct departure of the evolution from the expected rapid thermalization for isolated but interacting systems. Complementary to quenching the application of time-periodic driving is now emerging as a key tool for controlling many-body systems on microscopic time scales. In cold atom systems this is achieved by directly modulating the optical lattice potential~\cite{eckardt2017rmp,goldman2016nat}, while in condensed matter resonant THz excitation of low-energy structural and electronic degrees of freedom is opening up similar means of control~\cite{mankowski2016rep}. As such the possibility of stabilizing, enhancing and switching between various forms of order like superconductivity~\cite{hu2014nat,mitrano2016nat} and charge-density wave~\cite{singer2016prl} is a tantalizing prospect. Of particular fundamental and technological interest is the ultrafast control of magnetism~\cite{forst2011prb,caviglia2013prb,caviglia2012prl,forst2015nat,mikhaylovskiy2015nat,nova2016nat,baierl2016nat} that may have applications in magnetic storage devices~\cite{kirilyuk2010rmp,mentink2014prl,mentink2015nat}.

Motivated by these developments we examine the influence of strong periodic driving on the paradigmatic case of antiferromagnetic (AFM) N\'eel state relaxation within the repulsive Hubbard model. In the absence of driving the mechanism underpinning the melting of long-ranged antiferromagnetic order depends on the quench of the ratio of interactions to hopping $U/\nu_0$ from the atomic limit~\cite{balzer2015prx}. For $U \gg \nu_0$ local moments and their exchange coupling are retained during the evolution, but the quench results in the rapid nucleation of charge excitations whose motion on top of the spin background scrambles the staggered magnetization. For $\nu_0 \gtrsim U$ the melting is governed by the dynamics of residual quasi-particles that leads to the fast decay of both the local moments and long-range order. Suddenly introducing periodic driving is expected to have a profound influence on this relaxation.

Our focus is on the Hubbard model for a Bethe lattice in infinite dimensions where non-equilibrium dynamical mean-field theory (NE-DMFT) can solve for the dynamics. We study the relaxation process in three frequency $\omega$ regimes: (i) high-frequency $\omega \gg U,\nu_0$, (ii) resonant $l\omega = U > \nu_0$ with integer $l$, and (iii) in-gap off-resonant $U > \omega > \nu_0$. The high-frequency regime (i) leads to the well-known renormalization of the hopping $\nu_0$ and consequently the decay process for $\nu_0 > U$ to be switched from a quasi-particle to charge-excitation mechanism. Resonant driving (ii) and the resulting effective Hamiltonian describing the system have been the subject of recent theory work~\cite{bukov2016prl}. This indicates that interactions can be modified significantly, eliminated all together or mimic signatures of those of an opposite sign. This regime therefore allows for the reverse switch for $U > \nu_0$ from a charge-excitation to quasi-particle decay mechanism. For in-gap driving away from resonances (iii) we show that the increase of charge excitations is compensated by the suppression of their hopping, leaving the melting mechanism unaffected. We then discuss how by combining these driving regimes we can control the evolution of the system to induce a particular magnetic order parameter, and an ultrafast reversal of dynamics.

A significant contribution of this work is to demonstrate the accuracy and applicability of matrix product state techniques to the time-dependent Hamiltonian NE-DMFT impurity problem without increasing the number of bath sites compared to the time-independent case. This adds further impetus to the use of these methods for more complex problems involving other symmetry-broken states like superconductivity.

The plan of the paper is as follows. In \secr{sec:model} we introduce the driven Hubbard model to be considered, and briefly describe the key predictions of Floquet theory. In \secr{sec:dmft} we outline the framework of NE-DMFT, with particular emphasis on the Hamiltonian formulation of the impurity problem and the matrix product state technique we use as an impurity solver. Details of the DMFT setup for the N\'eel state melting are also described. In \secr{sec:undriven} we present results for the melting without driving, extending earlier work~\cite{balzer2015prx}, and providing a baseline for the expected behavior. This is followed in \secr{sec:driven} by the main results of this paper describing the effects of driving in the three regimes outlined. Finally we conclude in \secr{sec:conclusion}. 

\section{Model}
\label{sec:model}

\subsection{Driven Fermi-Hubbard lattice} \label{driven_hubbard}
We focus on the Fermi-Hubbard model at half-filling given by the Hamiltonian
\begin{equation}
H_{\rm Hub} = -J_0\sum_{\langle i,j\rangle,\sigma} c_{i\sigma}^{\dagger}c_{j\sigma}+U\sum_i\left(n_{i\uparrow}-\frac{1}{2}\right)\left(n_{i\downarrow}-\frac{1}{2}\right),
\end{equation}
where $c^\dagger_{i\sigma}$ creates an electron at site $i$ with spin $\sigma = \uparrow,\downarrow$, $n_{i\sigma} = c^\dagger_{i\sigma}c_{i\sigma}$ is the corresponding number operator, $J_0$ is the hopping amplitude between nearest-neighbor sites $\langle i,j\rangle$, and $U$ is the repulsive on-site interaction. For simplicity we consider a Bethe lattice in the limit of infinite coordination number $Z$ and hopping $J_0 = \nu_0/\sqrt{Z}$, where $\nu_0$ corresponds to the unit of energy. Physically this system can be envisaged as a cycle-free rooted tree in the $x$-$y$ plane with equidistant spacing $a$ between each site. Despite this it nonetheless mimics many properties expected of higher dimensional regular bipartite lattices and its $Z \rightarrow \infty$ limit is where our numerical approach, NE-DMFT, is exact \cite{eckstein2014rmp}.  

We take the system as being subjected to a uniform unpolarized AC electric field propagating in the $z$ direction resulting in a time-periodic linear potential emanating from the root of the tree\footnote{An essentially identical setup for a hypercubic lattice would be a uniform electric field polarized along the body-diagonal~\cite{mentink2015nat}.} via a driving term
\begin{equation} \label{driving}
H_{\rm drv}(t) = \sum_j e a E_0\sin(\omega t)\, s_j n_j, 
\end{equation}
where $e$ is the electronic charge, $E_0$ is the amplitude of the field, $\omega$ is the angular frequency of the drive, $s_j$ is the shell containing site $j$\footnote{There are $s_j$ steps from the root of the tree to site $j$.}, and $n_j=n_{j\uparrow}+n_{j\downarrow}$. Moving to the rotating-frame via the unitary transformation (we will take $\hbar = 1$ throughout)
\begin{equation}
U(t) = \exp\left[\ii \phi(t) \sum_j s_j n_j\right],
\end{equation}
where $\phi(t) = -(e a E_0/\omega)\cos(\omega t)$, we transform $H(t) = H_{\rm Hub} + H_{\rm drv}(t)$ into a rotating-frame Hamiltonian
\begin{equation}
H_{\rm rot}(t) = \ii \dot{U}(t) U^\dagger(t) + U(t) H(t) U^\dagger(t),
\end{equation}
where $\dot{U}(t)$ is the time derivative of $U(t)$. The explicit driving term is therefore eliminated making $H_{\rm rot}(t)$ equivalent to $H_{\rm Hub}$ with a time-dependent Peierls phase on the hopping amplitude as
\begin{equation} \label{hopping_time}
J(t) = J_0\exp\left[\ii A(s_i-s_j)\cos(\omega t)\right],
\end{equation}
where we define a dimensionless driving amplitude $A = e a E_0/\omega$. Nearest-neighbor hopping implies $(s_i-s_j) = \pm 1$, as it only occurs between neighboring shells.

\subsection{Floquet theory}
Since we are studying a periodically driven system we give a brief overview of Floquet formalism applied to the driven Hubbard lattice; see Refs.~\cite{bukov2015adv,kuwahara2016ann} for recent reviews. For closed quantum systems, Floquet's theorem establishes that there is a complete set of solutions of the time-dependent Schr\"odinger equation
\begin{equation} \label{schrodinger}
i\frac{d}{dt}|\psi(t)\rangle=H(t)|\psi(t)\rangle,\quad\text{with}\quad H(t+T)=H(t),
\end{equation}
and $T=2\pi/\omega$ the period of the time-dependent Hamiltonian $H(t)$, of the form
\begin{equation}
|\psi(t)\rangle=e^{-\ii \varepsilon_{\alpha}t}|\psi_{\alpha}(t)\rangle,\quad\text{with}\quad|\psi_{\alpha}(t)\rangle=|\psi_{\alpha}(t+T)\rangle,
\end{equation}
where the quasi-energies $\varepsilon_{\alpha}$ lay in the ``Brillouin zone" $-\omega/2<\varepsilon_{\alpha}\leq\omega/2$. This is equivalent to that of Bloch's theorem in time. Expanding the time-periodic function 
\begin{equation}
|\psi_{\alpha}(t)\rangle=\sum_{m}e^{-\ii m\omega t}|\psi_{\alpha,m}\rangle,
\end{equation}
in terms of the Floquet modes $|\psi_{\alpha,m}\rangle$, then \eqr{schrodinger} is reduced to the eigenvalue problem
\begin{equation} \label{eigenvalue_problem}
(\varepsilon_{\alpha}+m\omega)|\psi_{\alpha,m}\rangle=\sum_{m'}H_{m-m'}|\psi_{\alpha,m'}\rangle,
\end{equation}
in terms of the Fourier components of the Hamiltonian
\begin{equation}
H_m=\frac{1}{T}\int_0^Tdt\,e^{\ii m\omega t}H_{\rm rot}(t).
\end{equation}
For the Fermi-Hubbard lattice with time-dependent hopping given in \eqr{hopping_time}, this results in Fourier Hamiltonian terms~\cite{mentink2015nat}
\begin{align} \label{fourier_hami_terms}
\begin{split}
H_m=&-J_0\sum_{\langle ij\rangle,\sigma}(-1)^m\mathcal{J}_m\Big((s_i-s_j)A\Big)c_{i\sigma}^{\dagger}c_{j\sigma}\\
&+\delta_{m,0}U\sum_{j}n_{j\uparrow}n_{j\downarrow},
\end{split}
\end{align}
where the Coulomb repulsion is only present for the $m=0$ sector, and the hopping processes are renormalized by the $m$th Bessel function $\mathcal{J}_m\left((s_i-s_j)A\right)$. Truncating the number of Floquet sectors, or taking some specific limits, it is possible to solve the eigenvalue problem in \eqr{eigenvalue_problem}. For example, by moving to an extended Hilbert space and using second-order perturbation theory on the couplings between the $m=0$ sector and the rest, the exchange interaction $J_{\text{ex}}$ was found to be~\cite{mentink2015nat}
\begin{equation} \label{in_gap_floquet}
\frac{J_{\text{ex}}(A,\omega)}{J_{\text{ex}}(A=0)}=\sum_{n=-\infty}^{\infty}\frac{\mathcal{J}_{|n|}(A)^2}{1+n\omega/U},
\end{equation}
with $\nu_0/U\ll1$ so the exchange of the non-driven case $J_{\text{ex}}(A=0)$ can be well defined. In particular this result shows how high-frequency driving $\omega \gg U, \nu_0$ will lead to a reduction of $J_{\text{ex}}$. With in-gap driving $\nu_0 < \omega < U$, on the other hand, it is possible to increase $J_{\text{ex}}$, which can be exploited to enhance exchanging pairing below half-filling~\cite{jonathan2016}. For sufficiently strong driving $A$  \eqr{in_gap_floquet} even predicts that the sign of $J_{\text{ex}}$ can be flipped. At half-filling it is thus possible for periodic driving to induce an ultrafast reversal of the magnetic dynamics of a system~\cite{mentink2015nat}.

The application of high-frequency and strong-coupling expansions to the driven Hubbard lattice were recently placed on an equal footing~\cite{bukov2016prl} where both can be seen as a form of generalized Schrieffer-Wolf transformation. This formalism is particularly useful for resonant driving case $\nu_0\ll U=l\omega$, for integer $l$. There the effective Hamiltonian $H^{0}_{\rm eff} = H_0$ to zeroth-order in $1/\omega$, equal to the time-averaged rotating-frame Hamiltonian, is
\begin{equation} \label{bukov_hami}
H^{(0)}_{\text{eff}}=\sum_{\langle ij\rangle,\sigma}\left\{-J_{\text{eff}}g_{ij\sigma}-K_{\text{eff}}\left[(-1)^{l\eta_{ij}}h_{ij\sigma}^{\dagger}+\text{H.c.}\right]\right\},
\end{equation}
with the operators
\begin{subequations}
\begin{align}
&h_{ij\sigma}^{\dagger}=n_{i\bar{\sigma}}c_{i\sigma}^{\dagger}c_{j\sigma}(1-n_{j\bar{\sigma}}), \label{bukov_opers_h}\\
&g_{ij\sigma}=(1-n_{i\bar{\sigma}})c_{i\sigma}^{\dagger}c_{j\sigma}(1-n_{j\bar{\sigma}})+n_{i\bar{\sigma}}c_{i\sigma}^{\dagger}c_{j\sigma}n_{j\bar{\sigma}}, \label{bukov_opers_g}
\end{align}
\end{subequations}
where $\bar{\uparrow}=\downarrow$ and $\bar{\downarrow}=\uparrow$, and the parameters
\begin{equation} \label{bukov_params}
J_{\text{eff}}=J_0\mathcal{J}_0(A),\quad K_{\text{eff}}=J_0\mathcal{J}_l(A),
\end{equation}
with $\eta_{ij}=1$ for $i>j$ and $\eta_{ij}=0$ for $i<j$. The term $h^{\dagger}_{ij\sigma}$ corresponds to creation and annihilation of holon-doublon pairs between sites $i$ and $j$, and $g_{ij\sigma}$ represents holon-doublon and projected-fermion hopping. The weights of these processes can be manipulated by means of the driving amplitude $A$, as indicated in Eq.~\eqref{bukov_params}, leading to novel physics that we will illustrate in \secr{sec:driven}. The first-order correction in $1/\omega$ is $H^{(1)}_{\rm eff} = (1/\omega)\sum_{m=1}^\infty [H_m,H_{-m}]/m$ and gives rise to driving modified exchange terms.

\section{Non-equilibrium DMFT}
\label{sec:dmft}
To study the dynamics of the driven high-dimensional Fermi-Hubbard lattice described in Section~\ref{driven_hubbard}, we use the non-equilibrium extension~\cite{eckstein2014rmp} of DMFT~\cite{georges1996rmp}. In brief, this approach consists of replacing the correlated lattice by an impurity site, which retains the strictly local interactions of the lattice, and a coupling to an effective \textit{mean-field} $\Lambda_{\sigma}(t)$ representing the rest of the system. This mean-field must be determined self-consistently and is time-dependent in order to capture dynamical fluctuations arising from the exchange of particles with the impurity separated by a time $t$. For non-equilibrium configurations, which in general do not possess time-translational invariance, the dynamical mean-field depends on two times, and is denoted as $\Lambda_{\sigma}(t,t')$. Although this mapping is exact only in the limit of infinite dimensions (equivalent to $Z \rightarrow\infty$), it has been successfully used as the starting point of different approaches for finite-dimensional lattices including nonlocal spatial correlations~\cite{lichtenstein2000prb,maier2005rmp,rubtsov2009prb,loon2016prb,herrmann2016prb}. 

\subsection{Hamiltonian formulation} \label{hami_solver}
A variety of methods have been developed to solve the effective impurity problem of strongly correlated non-equilibrium high dimensional systems; see Ref.~\cite{eckstein2014rmp} for detailed descriptions. These include weak- and strong-coupling perturbation theories, which cannot deal with intermediate interaction regimes, and continuous-time quantum Monte-Carlo algorithms, that are limited by sign problems. To overcome these difficulties the use of a so-called \textit{Hamiltonian-based impurity solver}, which has been successful for equilibrium DMFT~\cite{georges1996rmp,hallberg2004prl}, has been recently proposed for the non-equilibrium case~\cite{gramsch2014prb}. The approach consists of discretizing the mean-field environment $\Lambda_{\sigma}(t,t')$ by mapping the effective impurity problem to a time-dependent single-impurity Anderson model (SIAM). This new problem corresponds to the impurity site coupled to an in-principle infinite, but in practise finite number of noninteracting bath orbitals $L_{\rm bath}$, described by the Hamiltonian
\begin{equation} \label{hami_siam}
H_{\text{SIAM}}(t)=H_{\text{imp}}+H_{\text{bath}}(t)+H_{\text{hyb}}(t),
\end{equation}     
with
\begin{subequations}
\begin{align}
&H_{\text{imp}}=U\left(n_{0\uparrow}-\frac{1}{2}\right)\left(n_{0\downarrow}-\frac{1}{2}\right)-\mu\sum_{\sigma}n_{0\sigma},\label{line1}\\
&H_{\text{hyb}}=\sum_{p>0,\sigma}\left(V_{0p}^{\sigma}(t)c_{0\sigma}^{\dagger}c_{p\sigma}+\text{H.c.}\right),\label{line2}\\
&H_{\text{bath}}=\sum_{p>0,\sigma}\left[\epsilon_{p\sigma}(t)-\mu\right]c_{p\sigma}^{\dagger}c_{p\sigma}.\label{line3}
\end{align}
\end{subequations}
Here we denote with $0$ the impurity site and bath orbitals by the index $p>0$, $\mu$ is the chemical potential, $V_{0p}^{\sigma}(t)$ is the time-dependent hopping amplitude of a fermion with spin $\sigma$ from orbital $p$ to the impurity, and $\epsilon_{p\sigma}(t)$ is the time-dependent on-site energy of bath orbital $p$ for spin $\sigma$. Equation~\eqref{line1} corresponds to the local impurity Hamiltonian possessing the repulsive interaction of the original lattice problem. Equation~\eqref{line2} describes the hybridization between the impurity and bath orbitals, and \eqr{line3} gives the on-site bath Hamiltonian. In this system there are only direct couplings between the bath orbitals and the impurity, therefore it is sketched in \figr{tebd}(a) as the impurity surrounded by bath orbitals in a star geometry.

The dynamics of the impurity site of the SIAM corresponds to that of each site of the original high-dimensional lattice once self-consistency conditions are met. Thus the original problem is solved by performing the time evolution of the SIAM, given the time-dependent parameters of \eqr{line1}-~\eqr{line3}, as obtained by the self-consistency to be outlined in \secr{melting_setup}. While simpler than the original lattice system, solving the impurity problem is still non-trivial. To aid this the use of state-of-the-art experimental quantum technologies has recently been proposed based on quantum simulating the dynamics of the SIAM with trapped ions~\cite{juha2016sci} and superconducting qubits~\cite{juha2016epj}. Otherwise popular approaches on a classical computer include exact diagonalization~\cite{gramsch2014prb}, multi-configurational time-dependent Hartree Fock~\cite{balzer2015prb} and matrix product state (MPS) calculations~\cite{alexander2014prb,balzer2015prx}. In this work we exploit a slightly different version of the latter as we now describe.  

\subsection{Matrix product impurity solver} \label{mps_solver}
\begin{figure}
\begin{center}
\includegraphics[scale=0.8]{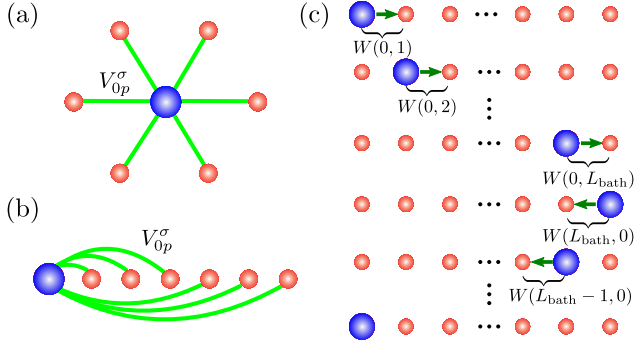}
\caption{\label{tebd} (a) Star geometry of the SIAM, with the impurity site (blue big circle) surrounded by noninteracting bath orbitals (small red circles). The green lines indicate the hybridization between both. (b) Linear chain geometry of the SIAM, with the impurity on the left edge and long-range hopping to the bath orbitals. (c) td-DMRG sweep for a single time step, where the impurity moves across the lattice by the action of swap gates.}
\end{center}
\end{figure}

The MPS tensor network has a one-dimensional chain geometry and as such this has made it an extremely successful ansatz for describing short-ranged interacting one-dimensional quantum systems. While the star geometry of the SIAM is straightforward with exact diagonalization, it does pose an issue for MPS approaches. To apply MPS we therefore reshape the system as a chain with the impurity site on an edge and long-range hopping, as depicted in \figr{tebd}(b). It is not \textit{a priori} clear whether a MPS simulation can be performed efficiently for this system, since entanglement is generally expected to grow very fast in the presence of long-range coupling. However it has been shown that entanglement grows slowly due to the inhomogeneous distribution of the couplings across the lattice, making the problem suitable for matrix product calculations~\cite{alexander2014prb}. For non-equilibrium problems, Refs.~\cite{alexander2014prb,balzer2015prx} successfully applied MPS simulation methods to the SIAM with long-range hopping using a Krylov-based time evolution algorithm~\cite{krylov2012njp}. 

In our work, we show that a conventional time-dependent density matrix renormalization group (td-DMRG) approach based on Trotterized two-site unitaries~\cite{vidal2004prl,schollwock2011ann} works similarly well, not only for time-independent evolution but also for periodic driving. The codes used here were implemented with the open-source Tensor Network Theory (TNT) library~\cite{tnt,tnt_review1}. The long-range interaction is dealt with as shown in \figr{tebd}(c). We perform the evolution during a single time step of length $\delta t$ by means of a second-order Trotter expansion
\begin{equation}
\exp\Big(-iH_{\text{SIAM}}\delta t\Big)\approx\left(\prod_{p=1}^{L_{\text{bath}}}W(0,p)\right)\left(\prod_{p=L_{\text{bath}}}^{1}W(p,0)\right),
\end{equation}
which consists of a left-to-right sweep of local two-site gates $W(0,p)$ followed by a right-to-left sweep of gates $W(p,0)$, defined by
\begin{subequations}
\begin{align}
W(0,p)=S\exp\Big(-\ii h(0,p)\frac{\delta t}{2}\Big),\\
W(p,0)=S\exp\Big(-\ii h(p,0)\frac{\delta t}{2}\Big).
\end{align}
\end{subequations}
Here $S$ is the fermionic swap gate that exchanges the position of the impurity with the nearest neighbor in the direction of the sweep, $h(0,p)$ is the SIAM Hamiltonian of impurity site $0$ and bath site $p$ for the left-to-right sweep where the impurity is to the left of the bath site, and similarly for $h(p,0)$ in the right-to-left sweep where the impurity is to the right of the bath site.

In this scheme the impurity is moved across the chain, becoming nearest neighbor of every bath orbital at some point in the sweep, and thus allowing for a standard implementation of each two-site gate. 
Our td-DMRG calculations were performed using particle-number conservation for both spins $\sigma=\uparrow,\downarrow$, bath size $L_{\text{bath}}=20$, and $U$-dependent time steps $5\times10^{-3}\leq\delta t\nu_0\leq5\times10^{-2}$. The decimation process during the time evolution was controlled by fixing a maximal truncation error per time step, with a maximal internal dimension of $\chi=1500$. The increase of $\chi$ during the evolution is determined by the Hamiltonian parameters, being fast for systems with strong correlations. This increase limits the final times that can be accessed within the algorithm, as the states of each step must be saved to perform the DMFT self consistency (see Sec.~\ref{melting_setup}) for which a huge amount of memory is required. However the timescales reached in our simulations, similar to those of the previous NE-DMFT+MPS works~\cite{alexander2014prb,balzer2015prx} and obtained with essentially the same computational effort, allow us to provide a complete discussion of the relevant physical processes underlying the dynamics. Approximate schemes to reach even longer times have been previously proposed~\cite{alexander2014prb}.

\subsection{Setup for N\'eel melting} \label{melting_setup}
Now we summarize the main steps to implement the NE-DMFT for an initial state that is a classical antiferromagnetic N\'eel state
\begin{equation}
|\Psi_{\rm N\'eel}\rangle = \prod_{i\in A} c^\dagger_{i\uparrow}\prod_{j\in B} c^\dagger_{j\downarrow} |0\rangle.
\end{equation}
A detailed description can be found in Refs.~\cite{gramsch2014prb,alexander2014prb,balzer2015prx}. The system consists of two homogeneous interpenetrating sub-lattices, $A$ for $\sigma=\uparrow$ and $B$ for $\sigma=\downarrow$. Since both lattices are identical except for their opposite magnetization, only the dynamics of one of them needs to be calculated, with that of the other sub-lattice following immediately. 

The central quantity of the NE-DMFT method and its self-consistency condition is the local single-particle Green function
\begin{equation} \label{green}
G_{k\sigma}(t,t')=-i\langle\mathcal{T}c_{k\sigma}(t)c_{k\sigma}^{\dagger}(t')\rangle,
\end{equation}
where $\mathcal{T}$ indicates time ordering along a Keldysh contour on which $t,t'$ are placed, and $c_{k\sigma}$ ($c_{k\sigma}^{\dagger}$) is the annihilation (creation) operator of a fermion at a particular site of sub-lattice $k=A,B$; since the sub-lattices are homogeneous, the site index is not included.

For a Bethe lattice its semi-elliptical density of states $D(\epsilon) = \sqrt{4\nu_0^2 - \epsilon^2}\,/(2\pi \nu_0^2)$ results in the lattice Green functions being related to the mean-fields $\Lambda_{k\sigma}(t,t')$ by a simple self-consistency condition,
\begin{equation} \label{self_consistency_initial}
\Lambda_{A(B),\sigma}(t,t')=\nu(t)G_{B(A),\sigma}(t,t')\nu^{*}(t'),
\end{equation}
where $J(t)=\nu(t)/\sqrt{Z}$ is the time-dependent hopping in the driven Hubbard model. Since the Green functions of the two sub-lattices are related by the symmetry condition $G_{A,\sigma}=G_{B,-\sigma}$, we may drop the $k$ index. Therefore \eqr{self_consistency_initial} becomes
\begin{equation} \label{self_consistency}
\Lambda_{\sigma}(t,t')=\nu(t)G_{-\sigma}(t,t')\nu^{*}(t').
\end{equation}
Additionally, the mean-field obtained from the SIAM with a star geometry is
\begin{equation} \label{mean_field_siam}
\Lambda_{\sigma}^{\text{SIAM}}(t,t')=\sum_pV_{0p}^{\sigma}(t)g_{p\sigma}(t,t')V_{p0}^{\sigma}(t'),
\end{equation}
with $g_{p\sigma}(t,t')$ the noninteracting Green function for the isolated bath orbital $p$, and whose exact analytical form is well known~\cite{eckstein2014rmp}. The time-dependent parameters of the SIAM, namely $V_{0p}^{\sigma}(t)$ and $\epsilon_{p\sigma}(t)$, must be chosen so
\begin{equation} \label{dmft_condition}
\Lambda_{\sigma}(t,t')=\Lambda_{\sigma}^{\text{SIAM}}(t,t').
\end{equation} 
If this condition is satisfied, the SIAM correctly captures the physics of the original high-dimensional lattice.

\begin{figure}
\begin{center}
\includegraphics[scale=0.85]{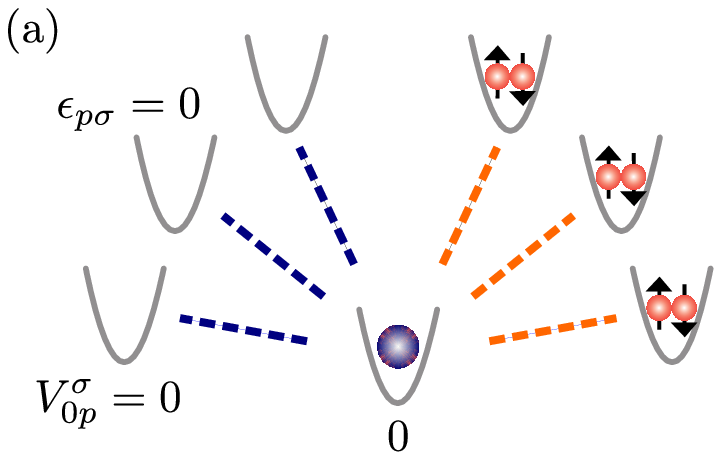}\\ \vspace{0.5cm}
\includegraphics[scale=0.9]{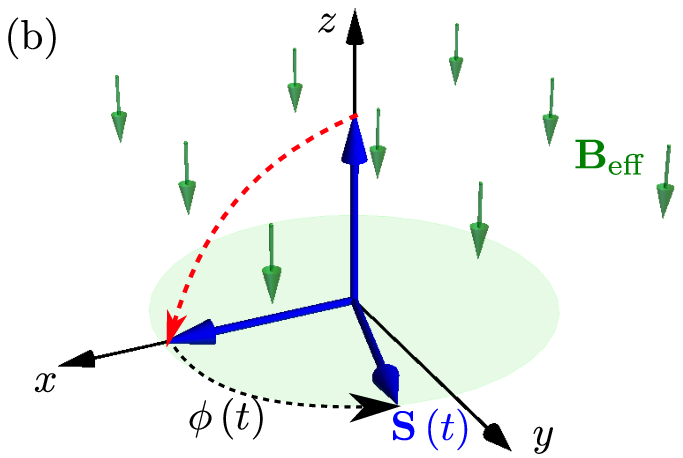}
\caption{\label{initial_states} (a) Initial state of the SIAM. The impurity site ($0$) is coupled to the bath orbitals, with $\epsilon_{p\sigma}=0$, through time-dependent hopping rates $V_{0p}^{\sigma}(t)$. At time $t=0$ the impurity is isolated, so $V_{0p}^{\sigma}(0)=0$. Initially half of the bath orbitals are empty and half are doubly occupied. The initial state of the impurity site, represented by a blue circle, depends on the particular problem of interest. (b) Scheme of spin precession dynamics. A single magnetic moment $\vec{S}(t)$ in the Bethe lattice is flipped to the $x$ direction, which precesses in the $x$-$y$ plane under the effective magnetic field $\vec{B}_{\text{eff}}$ (green arrows). From the precession angle $\phi(t)$ the exchange interaction $J_{\text{ex}}$ can be obtained as described in \secr{precession_section}.}
\end{center}
\end{figure}

Due to the particle-hole symmetry of the problem, and the possibility to freely choose $\epsilon_{p\sigma}(t=0)$, the initial state of the SIAM bath orbitals representing the high-dimensional lattice at half-filling acquires a simple form~\cite{gramsch2014prb}. Namely, half of the bath orbitals are empty, and the other half are doubly occupied, as sketched in \figr{initial_states}(a). The initial state of the impurity (site 0 of the SIAM) is determined by the initial configuration of the high-dimensional lattice. Taking the latter as a perfect N\'eel state, and calculating the dynamics only for the $k=A$ sub-lattice, the former corresponds to a single fermion with spin up. So the initial state of the SIAM is
\begin{equation}
|\psi_{\text{SIAM}}(t=0)\rangle=c^{\dagger}_{0\uparrow}\prod_{p=(L_{\text{bath}}/2)+1}^{L_{\text{bath}}}c_{p\uparrow}^{\dagger}c_{p\downarrow}^{\dagger}|0\rangle,
\end{equation}
with $|0\rangle$ the SIAM vacuum. The full implementation of the NE-DMFT requires several technical details and subtleties not mentioned here~\cite{gramsch2014prb,alexander2014prb,balzer2015prx}. Instead we sketch the main ingredients required for the algorithm. The basic steps for one of the sub-lattices are the following.
\begin{enumerate}

\item Start with a guess of the Green function $G_{\sigma}(t,t')$.

\item Obtain mean field $\Lambda_{\sigma}(t,t')$ from self consistency~\eqr{self_consistency}.

\item Calculate the SIAM hopping parameters $V_{0p}^{\sigma}(t)$ from DMFT conditions~\eqr{mean_field_siam} and~\eqr{dmft_condition}. This can be performed, for example, by means of a Cholesky decomposition~\cite{gramsch2014prb}. For simplicity, the on-site bath orbital energies are taken to be $\epsilon_{p\sigma}(t>0)=0$, so all time dependence is assigned to the hopping.

\item Obtain a new Green function $G_{\sigma}(t,t')$ by calculating \eqr{green} for the impurity site of the SIAM. The required time evolution of the initial state under Hamiltonian~\eqr{hami_siam} is calculated with td-DMRG as described in \secr{hami_solver}. 

\item Repeat steps 2-4 until convergence is reached.

\end{enumerate}
 
Once convergence has been obtained, single-site expectation values can be calculated. In particular we focus on the double occupation
\begin{equation}
d(t)=\langle n_{\uparrow}(t)n_{\downarrow}(t)\rangle,
\end{equation}
and the staggered magnetic order parameter $M_{\text{stagg}}(t) = \langle n_{\uparrow}(t)-n_{\downarrow}(t)\rangle$, along with
\begin{equation}
M(t)=\frac{M_{\text{stagg}}(t)}{1-2d(t)},
\end{equation}
which is normalized by the probability of the site being singly occupied~\cite{balzer2015prx}. The magnetization $M(t)$ signals the existence of antiferromagnetic order in the system, describing spin dynamics, and $d(t)$ indicates the formation of charge excitations during the dynamics.

\subsection{Spin precession dynamics} \label{precession_section}
To determine the AFM melting mechanism during the dynamics, the following setup was proposed in Ref.~\cite{balzer2015prx}. On a single probe site $o$ of sub-lattice A, the spin of the initially located fermion is flipped in the $x$ direction. This action leads to a change of $O(1/Z)$ on $\Lambda_{\sigma}$, and thus has a negligible back-action on the rest of the lattice~\cite{balzer2015prx}. The resulting dynamics corresponds to the local magnetic moment precessing in the effective mean field $\Lambda_{\sigma}(t,t')$ obtained in the DMFT simulations. So a new SIAM simulation is performed in which the spin at site $o$ (the impurity) starts in $x$ direction. The initial state of the SIAM is thus
\begin{equation}
|\psi_{\text{SIAM}}(t=0)\rangle=\frac{1}{\sqrt{2}}(c^{\dagger}_{0\uparrow}+c^{\dagger}_{0\downarrow})\prod_{p=(L_{\text{bath}}/2)+1}^{L_{\text{bath}}}c_{p\uparrow}^{\dagger}c_{p\downarrow}^{\dagger}|0\rangle,
\end{equation}
and the dynamics takes place with the same hopping parameters obtained during the DMFT. As the neighboring sites of the probe $o$ are oriented along the $z$ direction, we can assume that they lead to an effective parallel magnetic field $\vec{B}_{\text{eff}}=B_{\text{eff}}\hat{z}$. Under this approximation the dynamics is easily shown to correspond to a harmonic precession of the magnetic moment in the $x$-$y$ plane, with frequency $\Omega=\dot{\phi}(t)=B_{\text{eff}}$; this is depicted in \figr{initial_states}(b). Thus the components of the magnetic moment $\vec{S}(t)$ at the impurity site
\begin{subequations}
\begin{align}
S_x &= \frac{1}{2}\langle c_{0\uparrow}^{\dagger}c_{0\downarrow}+c_{0\downarrow}^{\dagger}c_{0\uparrow}\rangle, \\
S_y &= -\frac{i}{2}\langle c_{0\uparrow}^{\dagger}c_{0\downarrow}-c_{0\downarrow}^{\dagger}c_{0\uparrow}\rangle, \\
S_z &= \langle n_{0\uparrow}-n_{0\downarrow}\rangle,
\end{align}
\end{subequations}
with initial conditions $S_x(0)=1$ and $S_y(0)=0$, are described by the functions
\begin{equation} \label{circular_xy_traj}
S_x(t)=P_x\cos(\Omega_x t),\quad S_y(t)=P_y\sin(\Omega_y t),
\end{equation}
with amplitudes $P_x=P_y=1$ and frequencies $\Omega_x=\Omega_y=\Omega$. From these components an effective exchange interaction between the impurity and the mean field can be obtained, $J_{\text{ex}}=\dot{\phi}(t)/|M_{\text{stagg}}|$. Here
\begin{equation} \label{phi_angle}
\phi(t)=\tan^{-1}\left[\frac{S_y(t)}{S_x(t)}\right]
\end{equation}
is the angle of the spin in the $x$-$y$ plane~\cite{balzer2015prx} and $M_{\text{stagg}}$ is the staggered magnetization of the environment (obtained from the magnetic melting simulation), which defines the exchange interaction through $B_{\text{eff}}=J_{\text{ex}}M_{\text{stagg}}$. With this it is possible to find out whether the initial AF order is melted by (a) the decay of magnetic moments in time, (b) the decay of the effective exchange interaction while the moments persist, or (c) the energy exchange between charge and spin sectors (if (a) and (b) don't occur).  

\begin{figure}[t]
\begin{center}
\includegraphics*[width=0.8\columnwidth]{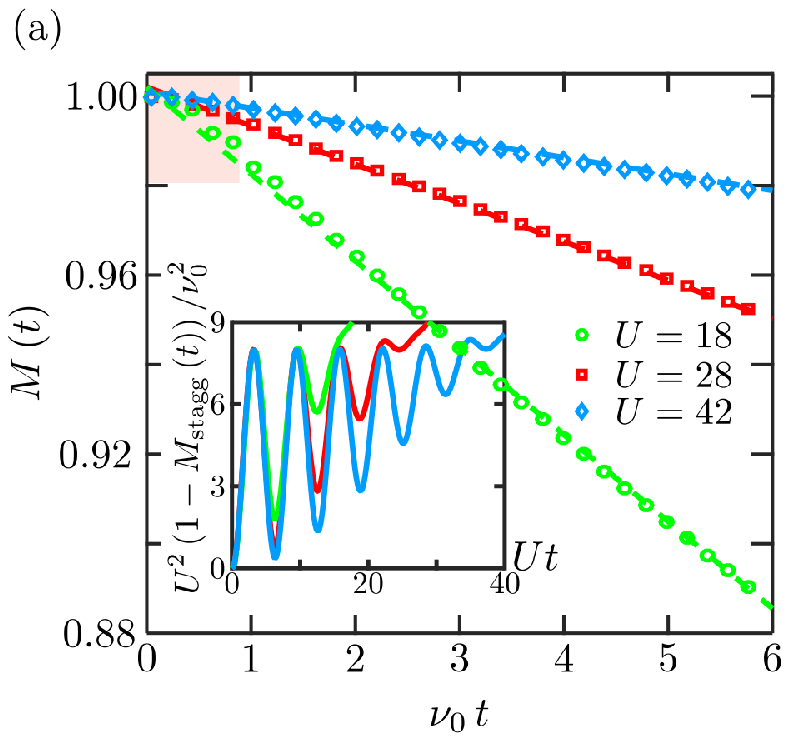}
\includegraphics*[width=0.8\columnwidth]{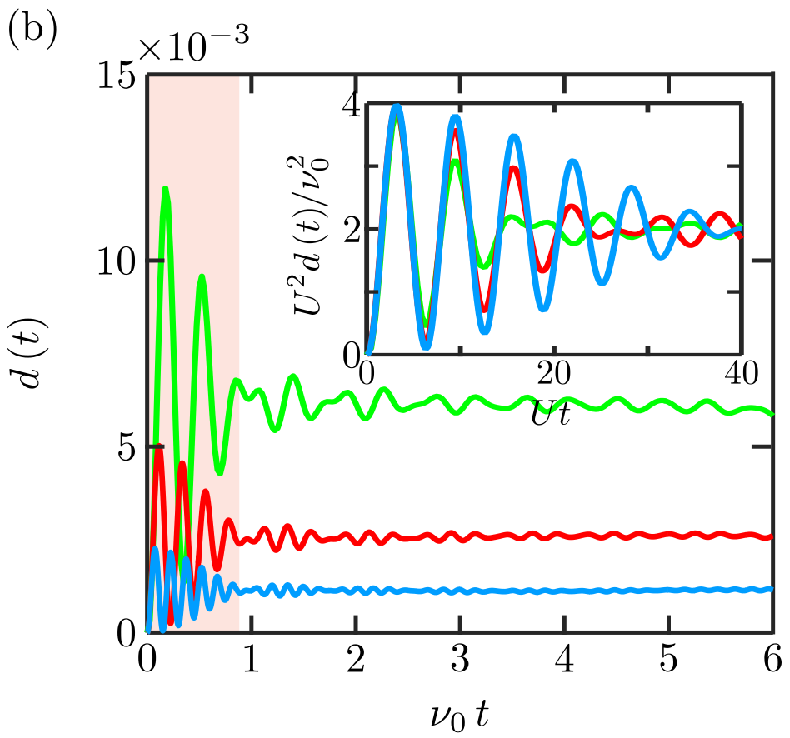}
\caption{\label{no_driving_M_d} Dynamics of a system with no driving and $U\gg1$. (a) Decay of order parameter $M(t)$. The symbols correspond to the results of the simulations, and the dashed lines to linear fits $M(t)=1-m\nu_0^3t/U^2$. For $U=18,28,42$ we have $m=6.29(2),6.77(1),6.49(1)$ respectively. Inset. Early-time dynamics of the staggered magnetization $M_{\text{stagg}}(t)$, with scaled axes. (b) Double occupation $d(t)$. Inset. Early-time dynamics, with scaled axes. The times for both insets correspond to the shaded areas of the main panels. We note that the oscillations at early times of both $M_{\text{stagg}}(t)$ and $d(t)$ have $U$-dependent decays, not captured by the perturbative results of Eqs.~\eqref{fit_M} and~\eqref{fit_d}. However we verified numerically that these are well described by a function $\sim\exp(-qU^2t^2)$ with constant $q$.}
\end{center}
\end{figure}

\section{Undriven magnetic melting}
\label{sec:undriven}
We start by discussing the physics of the undriven Fermi Hubbard model in a Bethe lattice ($A=0$ and $\nu=\nu_0=1$) to provide a baseline of expected effects. Even though this scenario has been studied in detail in Ref.~\cite{balzer2015prx} (see also Ref.~\cite{heidrich_meisner2015pra} for a study of the 1D case), we outline the melting process here focusing on the two limits of $U \gg \nu_0$ and $\nu_0 > U$. In \secr{sec:driven} we will describe how this physics is affected when the system is periodically driven.

\subsection{Melting via mobile charge excitations} \label{non_driven_large_U_section}
We first consider the case of strong Coulomb interactions $U \gg \nu_0$. In \figr{no_driving_M_d}(a) we show the decay in the magnetic order parameter $M(t)$, and correspondingly in \figr{no_driving_M_d}(b) the double occupation over the time $0\leq \nu_0 t \lesssim 6$ for different values of $U$. The behavior for both quantities divides up distinctly into times before and after $\nu_0t \approx 1$. For $\nu_0t > 1$ there is a linear decrease in $M(t)$ with a $U$ dependent gradient, while $d(t)$ converges to a $U$ dependent steady-state value.  Thus $M(t)$ and $d(t)$ relax at a different rate, the charge dynamics being faster than the spin dynamics. For $\nu_0 t < 1$ strong oscillations are observed in $M(t)$ and $d(t)$ that are well captured by a second-order time-dependent perturbation theory~\cite{heidrich_meisner2015pra} as
\begin{equation} \label{fit_M}
M_{\text{stagg}}(t)= 1 - \frac{8\nu_0^2}{U^2}\sin^2\left(\frac{Ut}{2}\right),
\end{equation}
and
\begin{equation} \label{fit_d}
d(t) = \frac{4\nu_0^2}{U^2}\sin^2\left(\frac{Ut}{2}\right).
\end{equation}
This is made clearer in the insets of \figr{no_driving_M_d}(a) and (b) where oscillations in $M(t)$ and $d(t)$ for different $U$'s collapse on top of one another and damp away within $\nu_0t \approx 1$. 

\begin{figure}[t]
\begin{center}
\includegraphics*[width=0.8\columnwidth]{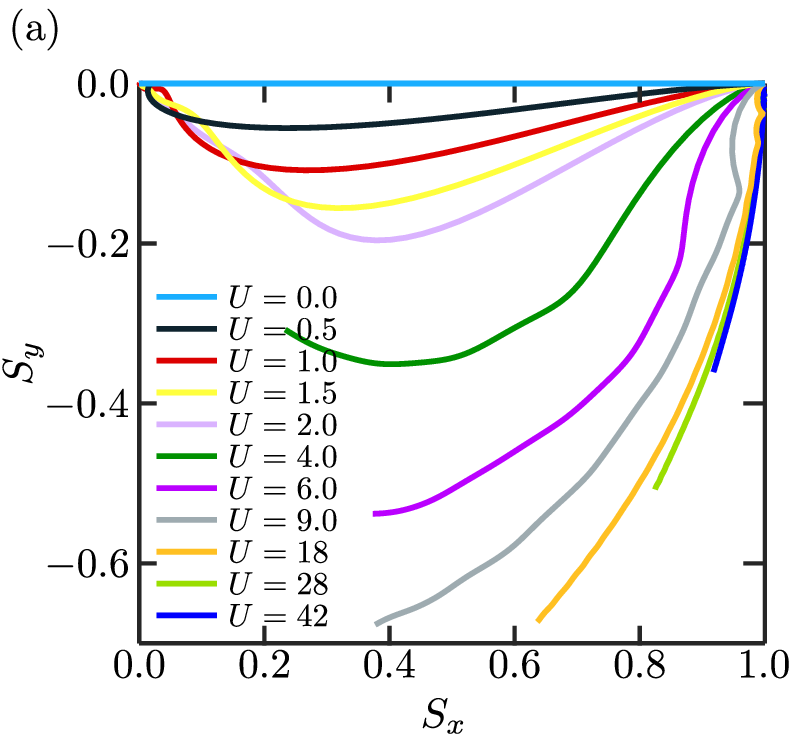}
\includegraphics*[width=0.8\columnwidth]{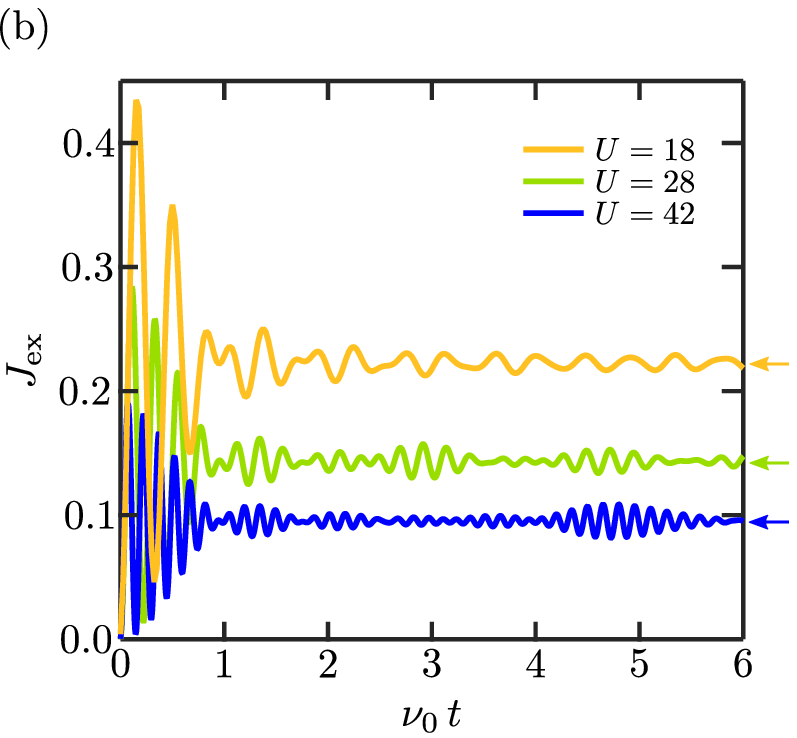}
\caption{\label{xy_no_driving} Spin precession dynamics for non-driven lattices. (a) Trajectory in the $x$-$y$ plane of the precessing magnetic moment. We checked numerically that this is approximately well described by Eq.~\eqref{circular_xy_traj}. (b) Exchange interaction $J_{\text{ex}}$ extracted from the dynamics of simulations with $U\gg1$. The perturbative exchange values~\eqref{pert_exchange} are indicated by arrows ($\leftarrow$). Matching colors among both panels correspond to equal values of $U$.}
\end{center}
\end{figure}

These results are consistent with the charge excitation decay mechanism outlined in Ref.~\cite{balzer2015prx}. The early time oscillations, resulting from the N\'eel state not being an eigenstate of the Hubbard model at finite $U$, nucleate a finite amount of charge excitations $d_{\rm ss} \approx 2\nu_0^2/U^2$ in the system, as highlighted in \figr{no_driving_M_d}(b). The motion of these excitations, with a hopping amplitude $\nu_0$, on top of a spin background then scrambles the AFM order. Owing to strong spin-charge coupling via the exchange interaction $J_{\rm ex}$ the kinetic energy of these charge carrier is transferred to the spin sector. This process is well captured by a so-called $t-J_z$ model~\cite{golez2014prb,balzer2015prx}, where it is found that for $\nu_0t > 1$ the number $f$ of spins flipped by a single carrier grows linear as $f(t) \sim 3\nu_0 t$ before saturation occurs. The decay of magnetization is therefore
\begin{equation} \label{M_decay_function_d}
M(t) \approx 1 - d_{\rm ss} f(t) = 1 - \frac{6 \nu_0^2}{U^2} (\nu_0 t).
\end{equation}
In \figr{no_driving_M_d}(a) we indeed find a linear decay with a gradient scaling as $1/U^2$ and coefficient close to $6$ for times $\nu_0 t> 1$.   

In this regime the spin precession dynamics, described in Section~\ref{precession_section}, gives rise to Fig.~\ref{xy_no_driving}. 
At early times $\nu_0 t < 1$ the exchange interaction behaves as $J_{\text{ex}}=2Ud(t)$, strongly oscillating around its static perturbative value
\begin{equation} \label{pert_exchange}
J_{\text{ex}}^{\text{pert}}=4\nu_0^2/U.
\end{equation}
For longer times $\nu_0 t>1$ both the precessing magnetic moment and the exchange interaction persist, confirming that the melting of $M(t)$ is not due to the suppression of local moments.

\subsection{Melting via residual quasi-particles}
\begin{figure}[t]
\begin{center}
\includegraphics*[width=0.8\columnwidth]{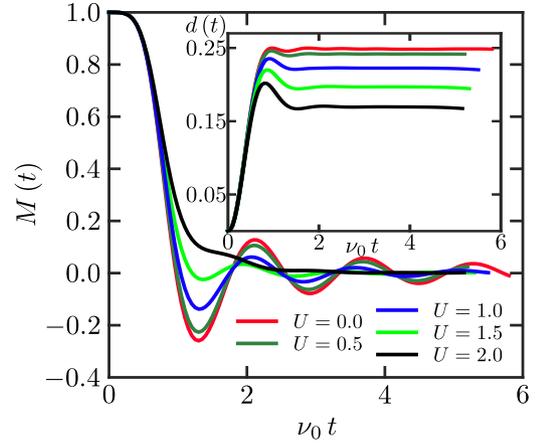}
\caption{\label{low_U_no_driving} Dynamics of a system with no driving and $U \lesssim 1$. Main panel. Order parameter $M(t)$. Inset. Double occupation $d(t)$. Matching colors correspond to equal values of $U$.}
\end{center}
\end{figure}

For $U/\nu_0 \lesssim 1$, close to the non-interacting integrable limit $U=0$, approximately conserved quasi-particles are expected to govern the dynamics leading to oscillatory behavior of $M(t)$ and prethermalization~\cite{balzer2015prx}. In Fig.~\ref{low_U_no_driving} we show the results for $M(t)$ and $d(t)$ for different low values of the Coulomb repulsion\footnote{Different final times are reached for each value of $U$ due to a different growth rate of $\chi$, as described in Sec.~\ref{mps_solver}. This is also observed in results to be presented in subsequent Sections.}. The magnetic order parameter oscillates and decays very fast to zero, and the double occupancy saturates to high values, approaching the non-interacting limit $d(t\rightarrow\infty)=0.25$ as $U$ decreases. The spin precession dynamics, shown in Fig.~\ref{xy_no_driving}, displays a very small decaying component in the $y$ direction and a rapidly decaying $x$ component. This indicates that the magnetic moments are short lived, and that an effective exchange interaction cannot be well defined. In this regime magnetic and charge dynamics occur on the same time scale, namely that of the hopping $\nu_0$. The melting of AFM order is therefore via a different mechanism to that seen at $U/\nu_0 \gg 1$, corresponding to the destruction of magnetic moments. 

\section{Magnetic melting in driven lattices}  
\label{sec:driven}
Having reviewed the behavior for the static case we now consider the impact of the driving, represented by the time-dependent hopping~\eqref{hopping_time}. To do so we explore in turn three different regimes for the driving frequency $\omega$: (i) high-frequency $\omega \gg U,\nu_0$, (ii) resonant $l\omega = U > \nu_0$ with integer $l$, and (iii) in-gap $U > \omega > \nu_0$. 

\subsection{High-frequency driving limit} \label{high_freq_section}

\begin{figure}[t] 
\begin{center}
  \includegraphics*[width=0.8\columnwidth]{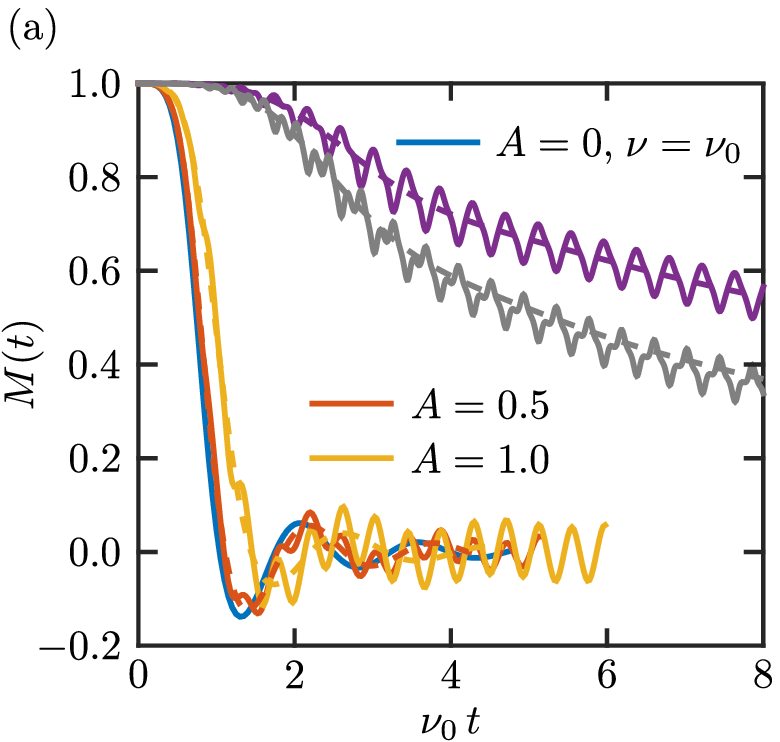} 
  \includegraphics*[width=0.8\columnwidth]{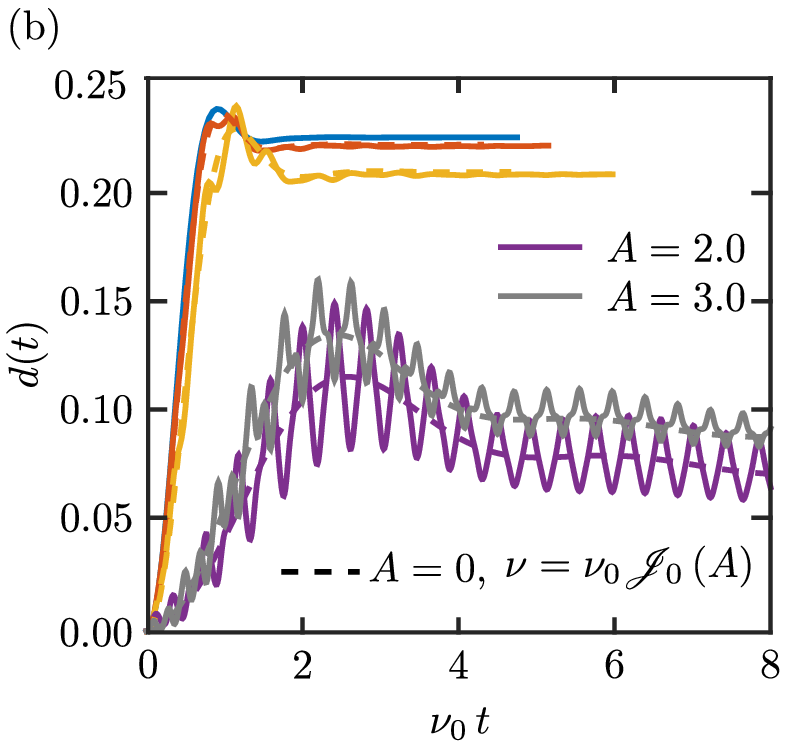} 
  \caption{Time evolution of a lattice with high-frequency driving. (a) Magnetic order parameter. (b) Double occupation. The solid lines correspond to systems with $\nu_0=U=1$, $\omega=15$ and different amplitudes $A$. The dashed lines correspond to systems with no driving, $U=1$ and renormalized hopping $\nu=\nu_0\mathcal{J}_0(A)$, with $A$ the driving amplitude of the solid line of the same color.}
  \label{high_frec_dynamics}
  \end{center}
\end{figure}  

The high-frequency limit, $\omega\gg U, \nu_0$, where the driving is much faster than all microscopic time scales of the system, is the conventional regime for Floquet theory. In Fig.~\ref{high_frec_dynamics} we show the behavior of $M(t)$ and $d(t)$ for $\nu_0=U=1$ and $\omega=15$. As expected for the weakly interacting regime, the static system displays magnetic order that decays very fast and features oscillations around $M(t) = 0$, while the double occupation quickly saturates to a value close to $d(t) = 0.25$, reproducing the results shown in Fig.~\ref{low_U_no_driving}. Low driving amplitudes $(A=0.5,1)$ have a weak impact on the relaxation. However, for larger amplitudes $A=2,3$ the dynamics is strongly affected, with a slower decay of $M(t)$ and a significant suppression of $d(t)$. This behavior is typical of static systems with a larger value of $U/\nu_0$.

\begin{figure}[t] 
\begin{center}
  \includegraphics*[width=0.8\columnwidth]{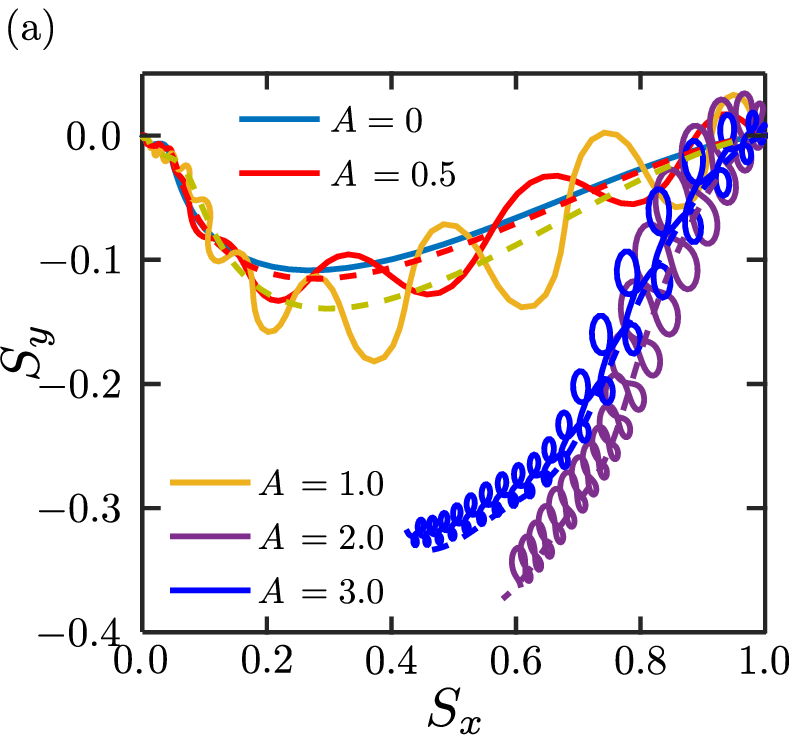} 
   \includegraphics*[width=0.8\columnwidth]{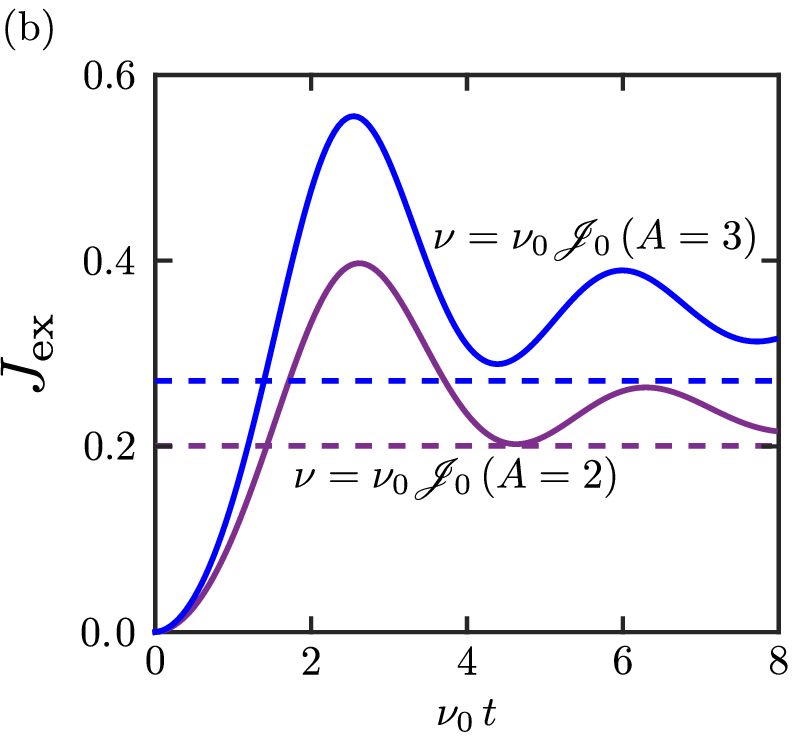} 
  \caption{Spin precession dynamics for high-frequency driving. (a) Trajectory in the $x$-$y$ plane of the precessing magnetic moment. The solid lines correspond to $\nu_0=U=1$, $\omega=15$ and different amplitudes $A$. The dashed lines correspond to systems with no driving, $U=1$ and hopping $\nu=\nu_0\mathcal{J}_0(A)$. (b) Exchange interaction $J_{\text{ex}}$ extracted from the precession dynamics of the system with renormalized hopping (solid lines), and perturbative values $4(\mathcal{J}_0(A)\nu_0)^2/U$ (dashed lines).}
  \label{high_frec_precession}
  \end{center}
\end{figure}

These results are well known from Floquet theory. In the high-frequency limit, different Floquet sectors are largely separated in energy, as suggested by Eq.~\eqref{eigenvalue_problem}. So the system can be effectively described by keeping only sector $m=0$. From Eq.~\eqref{fourier_hami_terms}, this corresponds to a Fermi-Hubbard model with the hopping being renormalized by a Bessel function $\mathcal{J}_0(A)$. Since $|\mathcal{J}_0(A)|<1$, this corresponds to a suppression of the hopping, or equivalently to an effective enhancement of $U/\nu_0$. 

As depicted in Fig.~\ref{high_frec_dynamics} for both $M(t)$ and $d(t)$, the dynamics of the driven system is (after averaging out small modulations at frequency $\omega$) very well captured by a static system with a hopping renormalized by $\mathcal{J}_0(A)$. This picture is confirmed by the spin precession dynamics; the results are shown in Fig.~\ref{high_frec_precession}. For weak driving amplitudes the precessing magnetic moment decays quickly, while for large amplitudes it persists for long times. The average dynamics is again well described by a non-driven system with a renormalized hopping along with an exchange interaction (obtained as described in Section~\ref{precession_section}) close to the renormalized perturbative value $4(\mathcal{J}_0(A)\nu_0)^2/U$. 

Our simulations thus corroborate the prediction from Floquet theory of effective hopping renormalization at high frequency driving. Choosing amplitudes $A$ so $|\mathcal{J}_0(A)|\ll1$ it is then possible to switch a system with $U/\nu_0 < 1$ from displaying quasi-particle melting to charge-excitation melting. Furthermore, by selecting $A$ so $\mathcal{J}_0(A)=0$, the hopping is completely suppressed (effectively corresponding to infinite Coulomb interaction) and the dynamics is frozen\footnote{This is exemplified in Section~\ref{combine_schemes} for a combined driving scheme to stabilize the magnetic order parameter or the number of charge excitations to a particular value.}. This phenomenon of \textit{dynamical localization} by coherent driving was predicted long ago in a noninteracting system with the same type of driving used here~\cite{dunlap1986prb}. It has also been studied theoretically in spin chains~\cite{das2010prb,roy2015pra}, and was observed experimentally in a system of cold atoms where a potential of the form of Eq.~\eqref{driving} was created by shaking an underlying optical lattice~\cite{lignier2007prl}.

\subsection{Resonant driving} \label{resonant_section}

\begin{figure}[t] 
\begin{center}
  \includegraphics*[width=0.8\columnwidth]{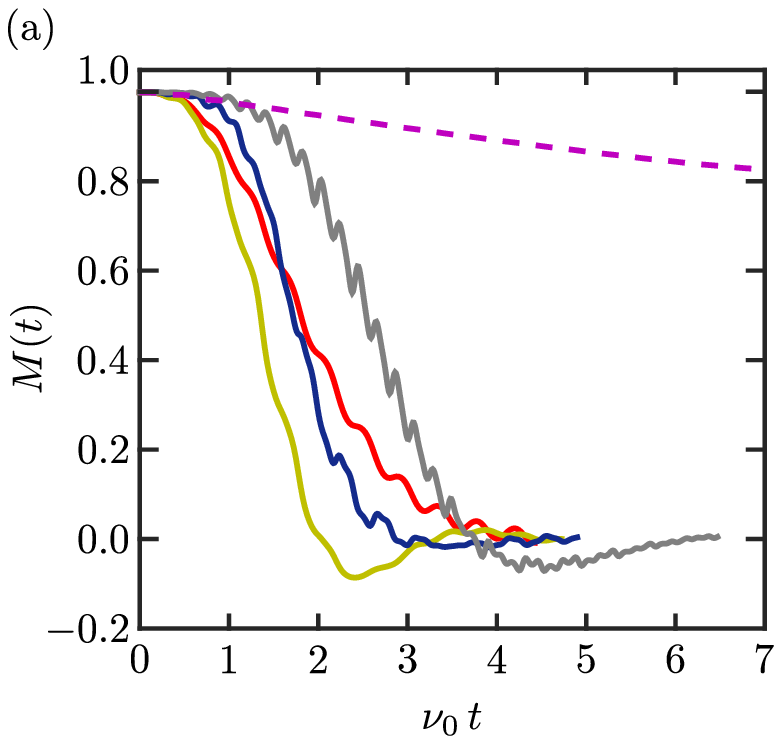} 
   \includegraphics*[width=0.8\columnwidth]{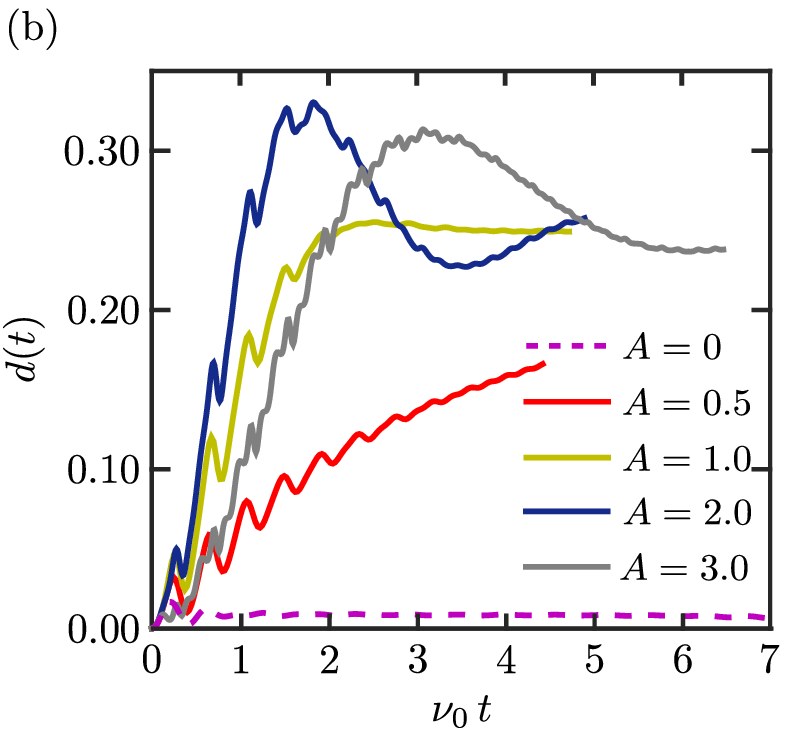} 
  \caption{Time evolution of a lattice with resonant driving. (a) Magnetic order parameter. (b) Double occupation. The solid lines correspond to $\nu_0=1$, $U=\omega=15$ and various amplitudes $A$. The dashed line corresponds to the non-driven case.}
  \label{resonant_driving}
  \end{center}
\end{figure}

We now decrease $\omega$ to the point that it is directly resonant as $\omega=U\gg\nu_0$ ($l=1$). The resulting dynamics is depicted in Fig.~\ref{resonant_driving} for $\omega=U=15$. In the non-driven case, the magnetic order decays very slowly and the double occupation remains low; the Coulomb repulsion is so large that fermions are prevented from hopping across the lattice. This picture is strongly modified when driving at resonance, even at weak amplitudes. The magnetic order decays very fast, and the double occupation increases to large values. Importantly, for $A=0.5$ the magnetization $M(t)$ has been suppressed at short times while $d(t)$ has not saturated. For strong driving $A > 0.5$ the magnetic order melting is still fast, and the charge dynamics occurs on a commensurate timescale. This shows that with resonant driving it is possible to control the speed of charge dynamics compared to spin dynamics up to the point of making the former slower, a behavior not seen in the non-driven case. 

\begin{figure}[t] 
\begin{center}
  \includegraphics*[width=0.8\columnwidth]{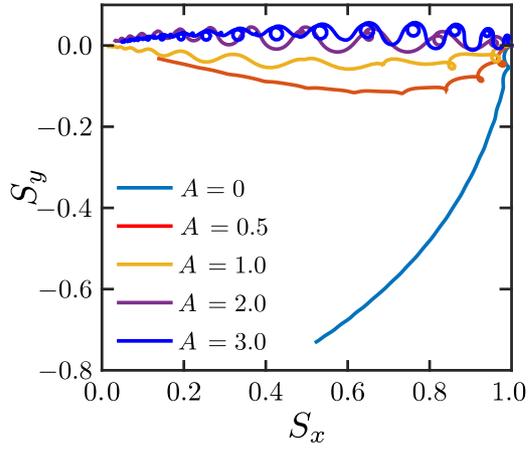} 
  \caption{Precession dynamics in the $x$-$y$ plane for $\nu_0=1$, resonant driving $l=1$ $(U=\omega=15)$ and various amplitudes $A$.}
  \label{resonant_driving_prec}
\end{center}  
\end{figure}

The magnetic melting mechanism is elucidated from the spin precession dynamics shown in the main panel of Fig.~\ref{resonant_driving_prec}. In the static case the magnetic moment is preserved for a long time, while for the resonantly driven case the magnetic moments decay very fast. Resonant absorption of energy from the drive thus destroys the local moments so an exchange interaction cannot be defined, and suppresses magnetic order. The behavior is extremely reminiscent of the static case with $U/\nu_0 < 1$, and suggests that resonant driving can be used to switch from a charge-excitation mechanism to quasi-particle melting. 

\begin{figure}[t] 
\begin{center}
  \includegraphics*[width=0.8\columnwidth]{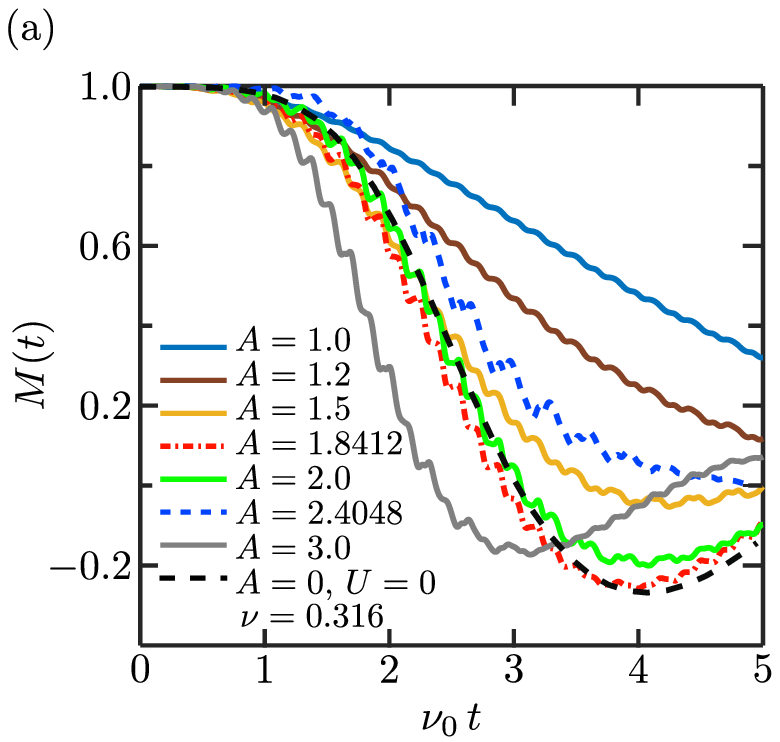}
  \includegraphics*[width=0.8\columnwidth]{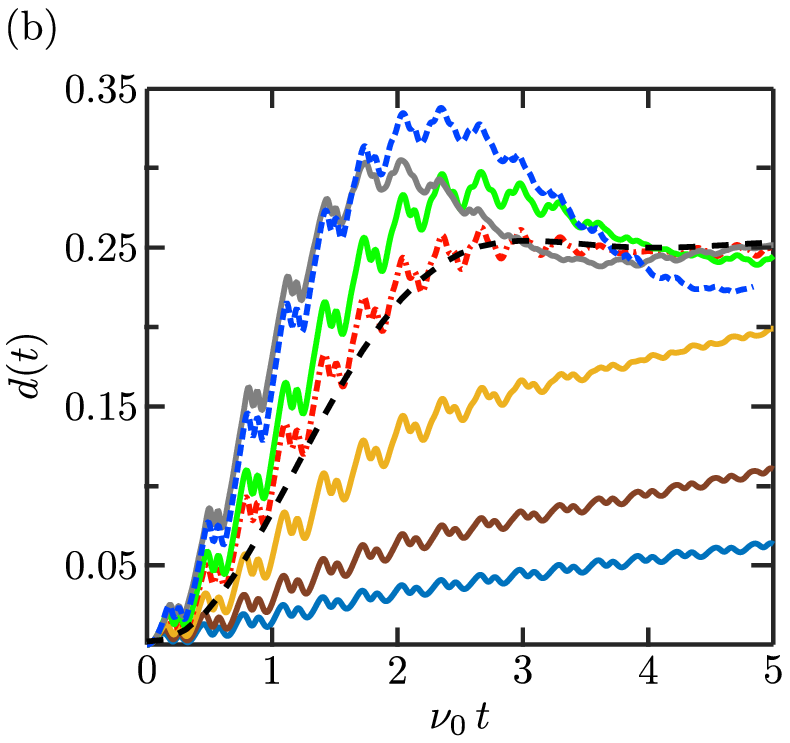} 
  \caption{Resonant $l=2$ dynamics for $U=40$, $\omega=20$ and several driving amplitudes $A$. (a) Magnetic order parameter. (b) Double occupation. We go from repulsive interactions ($A=1,1.2,1.5$) to an effective non-interacting case ($A=1.8412$, well described by a non-interacting lattice with renormalized hopping $\nu=\nu_0\mathcal{J}_0(A)=0.316$), to positive effective interactions ($A=2$). After creation of holon-doublon pairs is maximized ($A=2.4048$), the interactions start decreasing while still being positive ($A=3$).}
  \label{bukov_resonance_noninteracting}
  \end{center}
\end{figure}

Resonant driving with $l$ even can in fact map the interacting system exactly to the $U=0$ limit. As predicted in Ref.~\cite{bukov2016prl}, when the amplitude is such that $J_{\text{eff}}=K_{\text{eff}}$ in Hamiltonian~\eqref{bukov_hami}, the driving precisely induces holon-doublon creation terms that would exist if $U=0$ with no driving. We consider the situation $l=2$ ($\omega=U/2$) and $A=1.8412$ in Fig.~\ref{bukov_resonance_noninteracting}. The driven lattice is seen to be essentially equivalent to a non-interacting one with renormalized hopping $\nu_0\mathcal{J}_0(A)$ confirming this novel effect for an infinite-dimensional system. Therefore at this special driving strength and frequency $\omega = U/2$ we have induced an effective suppression of the Coulomb repulsion $U$, which corresponds to the opposite effect of the effective interaction enhancement at high-frequency driving.  

\subsubsection{Effective attractive interactions and pairing}
For larger driving strengths $A = 2,3$ at both the $l=1$ and $l=2$ resonances the double occupation shows an intriguing feature. In Fig.~\ref{resonant_driving} and Fig.~\ref{bukov_resonance_noninteracting} we see that $d(t) \approx 0.33$ at early times, and so exceeds the $d_{\rm ss} = 0.25$ value of the non-interacting limit~\cite{balzer2015prx,heidrich_meisner2015pra}. A similar effect has been reported before~\cite{tsuji2011prl} under very different conditions of a high-frequency drive, with large driving amplitudes $A$ so that $\mathcal{J}_0(A)<0$, flipping the sign of the hopping. There the appearance of $d(t) > 0.25$ was argued to imply that fermions are clustering due to an effective {\em attractive} interaction. Here similar signatures are seen, but crucially they occur on a much faster time scale and emerge from an initial state with zero double occupancies.   

\begin{figure}[t] 
\begin{center}
  \includegraphics*[width=0.8\columnwidth]{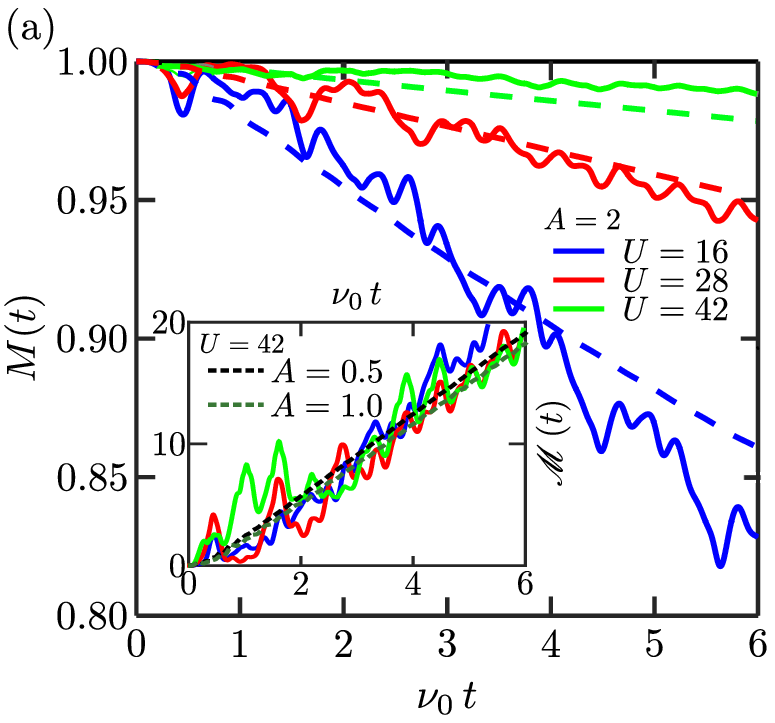} 
  \includegraphics*[width=0.8\columnwidth]{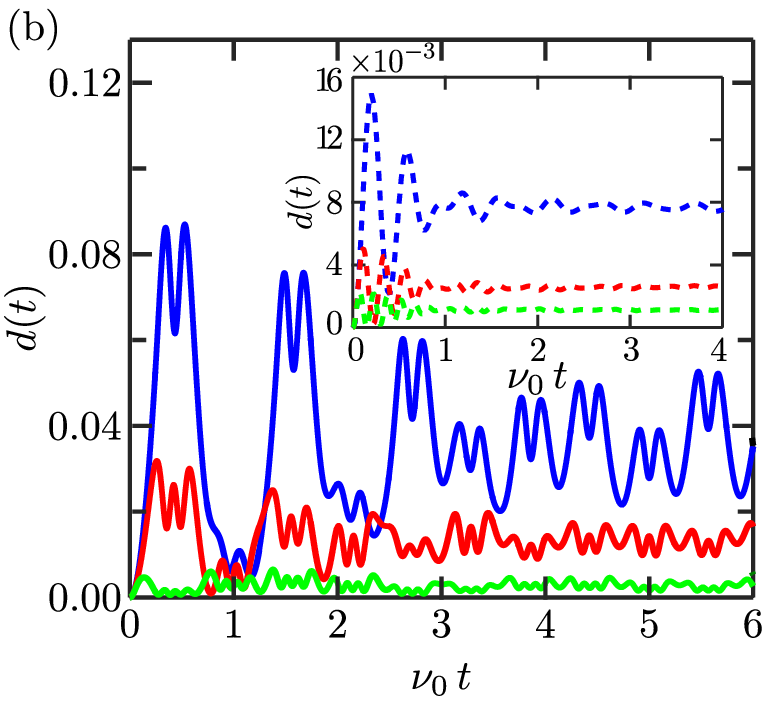} 
  \caption{Time evolution of a lattice with in-gap driving. The solid lines correspond to $\nu_0=1$, $A=2$, $\omega=11$ and different values of $U>\omega$. The dashed lines of the same color correspond to equal values of $U$ with no driving. (a) Magnetic order parameter. Inset. Collapse of $\mathcal{M}(t)=(1-M(t))/\mathcal{J}_0(A)\langle d\rangle$ as a function of time for the values of $U$ of the main panel; results for $U=42$ and different amplitudes $A$ have also been included, to further test Eq.~\eqref{melting_in_gap_equation}. (b) Double occupation. The results for non-driven systems are shown in the inset for clarity.}
  \label{ingap_dynamics}
  \end{center}
\end{figure} 

This can be qualitatively understood from the resonant effective Hamiltonian~\eqref{bukov_hami} with $l=1$. For $A=0.5,1$ we have that $J_{\text{eff}}>K_{\text{eff}}$, so the dynamics is dominated by holon and doublon hopping processes. On the other hand, for $A=2,3$ then $|K_{\text{eff}}|>|J_{\text{eff}}|$, so holon-doublon creation processes dominate, explaining the greater propensity for double occupancies. For $l=2$ and $A=2.4048$ we have that $J_{\text{eff}}=0$, so the dynamics is then entirely governed by creation and annihilation of doublon-holon pairs. This temporarily maximizes $d(t)$, as depicted in Fig.~\ref{bukov_resonance_noninteracting}(b). However in either case the effective Hamiltonian is far from simple, owing to frustration effects from overlapping holon-doublon creation terms. We therefore postpone a more detailed analysis of potential pairing~\cite{jonathan2016} to future work where it maybe of relevance to observations of light-induced superconductivity~\cite{hu2014nat,mitrano2016nat}.

\subsection{In-gap off-resonant driving} \label{ingap_section}

Now we consider the regime of in-gap driving $\nu_0<\omega<U$. For our analysis we fix $\omega$ and take different values of $U$, staying away from resonant points $U=l\omega$. In this form the divergences of $J_{\text{ex}}(A,\omega)$ predicted by Eq.~\eqref{in_gap_floquet} are avoided. The magnetic order parameter $M(t)$ and the double occupation $d(t)$ are depicted in Fig.~\ref{ingap_dynamics}, for both the driven and the non-driven cases. Although each time evolution shows a general decay of $M(t)$ with strong fluctuations, it closely follows the melting of the corresponding non-driven case. The double occupation, on the other hand, notably increases on average due to the driving, although still remains low.

\begin{figure}[t] 
\begin{center}
  \includegraphics*[width=0.8\columnwidth]{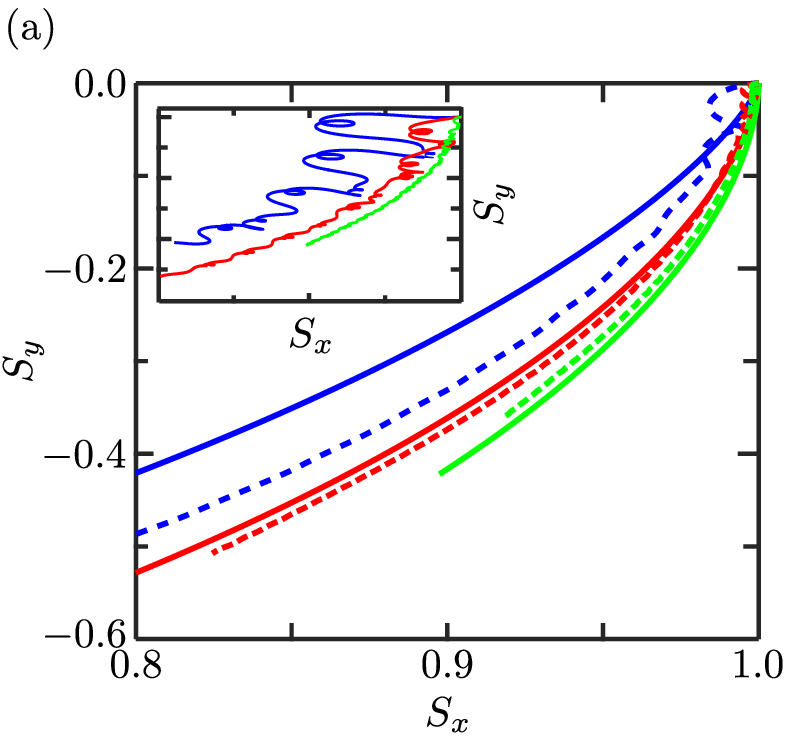} 
  \includegraphics*[width=0.8\columnwidth]{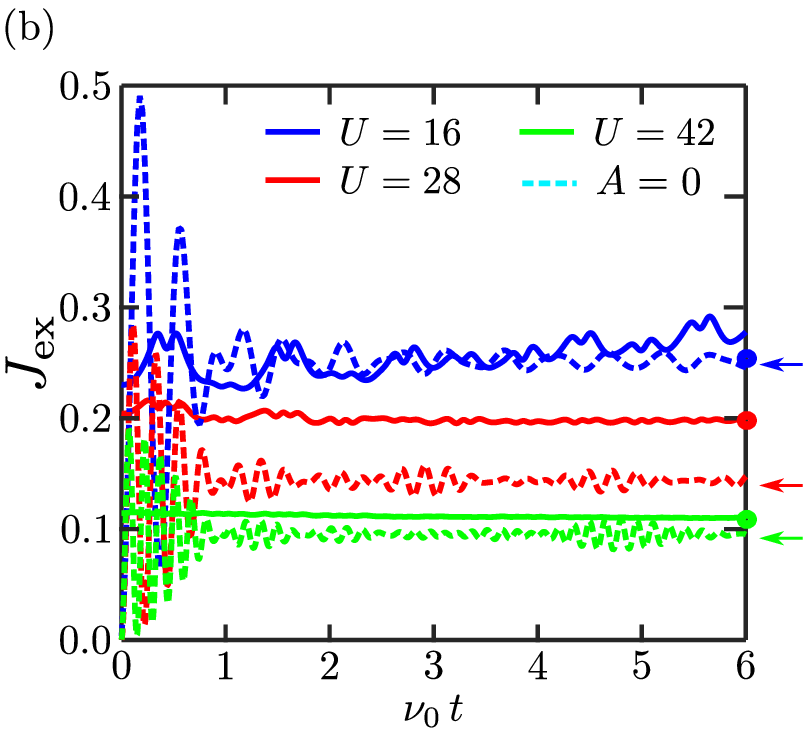} 
  \caption{(a) Trajectories in the $x$-$y$ plane of precessing magnetic moments in systems with $\nu_0=1$, $A=2$, $\omega=11$ and different values of $U>\omega$. The solid lines are the resulting trajectories when fitting $S_x$ and $S_y$ to the functions~\eqref{circular_xy_traj}, with the original trajectories being shown in the inset (with the same scale for both $S_x$ and $S_y$). The dashed lines of matching colors are for systems with equal $U$ and no driving. (b) Exchange interaction $J_{\text{ex}}$ for the driven dynamics extracted from the fitting functions~\eqref{circular_xy_traj} (solid lines; extracting $J_{\text{ex}}$ from the trajectories depicted in the inset of (a) is complicated due to their many loops), and for systems with equal $U$ and no driving (dashed lines). The perturbative exchange~\eqref{pert_exchange} for the non-driven cases are indicated by arrows ($\leftarrow$), and those from Floquet theory for in-gap off-resonant driving~\eqref{in_gap_floquet} by solid circles ($\bullet$).}
  \label{ingap_precession}
  \end{center}
\end{figure}

These results suggest that the melting mechanism of the off-resonant in-gap driving discussed here is the same of the strongly-interacting non-driven case, described in Sec.~\ref{non_driven_large_U_section}. In fact, as depicted in the inset of Fig.~\ref{ingap_dynamics}(a), besides oscillations the magnetic order parameter satisfies
\begin{equation} \label{melting_in_gap_equation}
M(t)\sim1-\langle d\rangle\tilde{f}(t),
\end{equation} 
with $\tilde{f}(t)=3\mathcal{J}_0(A)\nu_0t$ and $\langle d\rangle$ the time-average of $d(t)$. This behavior is completely analogous to that of Eq.~\eqref{M_decay_function_d} for the non-driven case, with renormalized hopping $\nu_0\mathcal{J}_0(A)$ as expected for in-gap off-resonant driving~\cite{jonathan2016}. So the magnetic melting emerges from the strong spin-charge coupling, being mostly dependent on the hopping and the number of charge excitations nucleated at early times, and weakly dependent on the value of the exchange interaction. The fact that the melting of the order parameter closely follows that of the non-driven case indicates that the effective suppression of the hopping by $\mathcal{J}_0(A)$ is approximately compensated by the increase of the double occupation.

This picture is reinforced by the spin precession dynamics. In Fig.~\ref{ingap_precession}(a) we show fitted trajectories of the precessing spin in the $x$-$y$ plane, which for each value of $U$ are very similar to those of the non-driven case. The magnetic moments thus remain long-lived under in-gap off-resonant driving. The resulting effective exchange interactions of the driven and non-driven cases are also similar, as shown in Fig.~\ref{ingap_precession}(b). The relation between both is very well captured by the Floquet result Eq.~\eqref{in_gap_floquet}. For the different $U$ we obtain that $1.02<J_{\text{ex}}(A,\omega)/J_{\text{ex}}(A=0)<1.40$, so the exchange interactions slightly increase with the driving. As the magnetic moments and their exchange are long-lived, the magnetic melting occurs due to energy transfer from charge to spin sectors.

In summary the results of this Section show that for in-gap driving away from resonances, a similar magnetic dynamics to that of the non-driven case is induced. The notably larger double occupancies created by the driving are compensated by the suppression of the hopping. In addition the exchange interaction is weakly modified. This approximately leads to the same magnetic melting rate in both the driven and non-driven scenarios, leaving the underlying melting mechanism unaffected.
 
\subsection{Dynamic control of magnetic melting}  \label{combine_schemes}

\begin{figure}[t] 
\begin{center}
\includegraphics*[width=0.8\columnwidth]{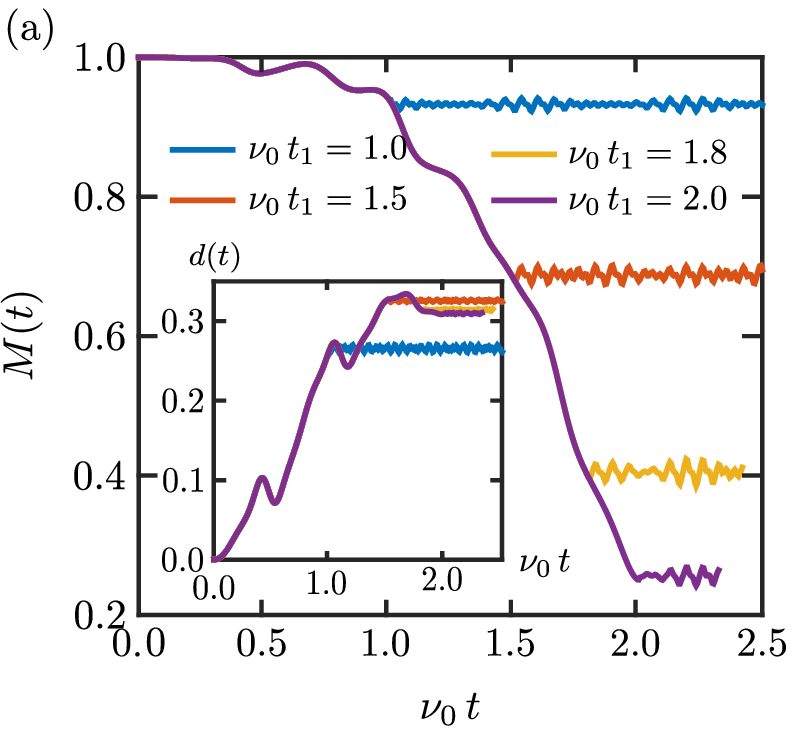}  
\includegraphics*[width=0.8\columnwidth]{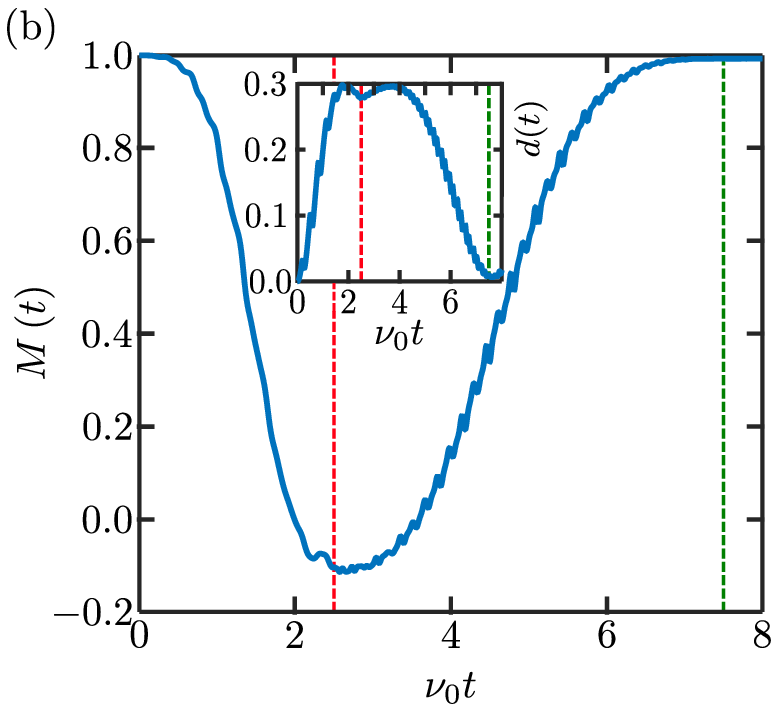}  
  \caption{(a) Simple protocol to stabilize a strongly-interacting system ($\nu_0=1$, $U=10$) with a particular magnetization (main panel) or double occupation (inset). From $t=0$ to $t=t_1$ resonant driving is applied ($A=2$, $\omega=10$) to quickly decrease the magnetization. Then for $t>t_1$ high-frequency driving is applied ($A=2.4048$ so $\mathcal{J}_0(A)=0$, $\omega=95$), which freezes the system in the final state of the first driving scheme. (b) Protocol to induce dynamics reversal on a strongly-interacting system at $l=1$ resonance ($U=\omega=20$), indicated by both the magnetic order parameter (main panel) and the double occupation (inset). From $t=0$ to $t_1=2.5$ (red dashed lines) the system in driven with $A_1=1.40$, and from $t_1=2.5$ onwards it is driven with $A_2=4.65$. In this form $\mathcal{J}_0(A_1)/\mathcal{J}_0(A_2)=-2.003$ and $\mathcal{J}_1(A_1)/\mathcal{J}_1(A_2)=-2.021$. Reversal of dynamics is observed at time $\nu_0 t\approx7.5$ (green dashed lines).}
  \label{combined_schemes}
  \end{center}
\end{figure}  

Finally we show two examples of how combining the different driving schemes previously described, it is possible to engineer a ultra-fast protocol to manipulate the magnetic order and pairing of a fermionic interacting system. The first corresponds to rapidly stabilizing a particular magnetization value from an initial N\'eel state with large $U/\nu_0$. An immediate form to do it is following a driving scheme like that depicted in Fig.~\ref{combined_schemes}(a). This consists of two stages. First the desired magnetization is quickly reached by resonant driving. Then the dynamics is frozen by high-frequency driving, maintaining the magnetization for the required time. This protocol could also be used for stabilizing a state with enhanced pairing, targeting a particular value of the double occupation instead of magnetization (inset of Fig.~\ref{combined_schemes}(a)).   

The second example illustrates the induction of dynamics reversal by an appropriate choice of the driving. In Ref.~\cite{mentink2015nat} a scheme for achieving this effect in long time scales $\nu_0 t\approx100$ was discussed. There a 1D system was allowed to evolve freely for some time, after which it was in-gap and non-resonantly driven in such a way that, from Eq.~\eqref{in_gap_floquet}, $J_{\text{ex}}(A,\omega)/J_{\text{ex}}(A=0)\approx-1$. By slowly ramping the driving field electronic excitations were strongly impeded, leading to spin-dominated dynamics which could be reversed by changing the sign of the exchange interaction.

Here we use a similar idea to induce a dynamics reversal on much shorter time scales $\nu_0 t<8$, but with a sudden quench of the amplitude of a $l=1$ resonant driving designed to invert the sign of the effective Hamiltonian~\eqref{bukov_hami}. The scheme consists of two stages, starting from a N\'eel state, and leading to the results depicted in Fig.~\ref{combined_schemes}(b). During the first stage the system evolves under resonant excitation with amplitude $A_1$ till time $t_1$, leading to a fast melting of the magnetic order and large charge excitations. For the second stage the driving amplitude $A_2$ is chosen so the sign of both $J_{\text{eff}}$ and $K_{\text{eff}}$ is inverted, while keeping their ratio to the values of the first stage approximately constant. The amplitudes used in Fig.~\ref{combined_schemes}(b) are such that $J_{\text{eff}}(A_1)/J_{\text{eff}}(A_2)\approx K_{\text{eff}}(A_1)/K_{\text{eff}}(A_2)\approx-2$. The dynamics for $t>t_1$ is thus slower than that for $t<t_1$ by a factor of $2$, in addition to the change of sign, resulting in an almost-complete reversal of both the magnetic order parameter and the double occupation after an evolution of length $2t_1$. The reversal is not perfect due to the higher-order terms in Hamiltonian~\eqref{bukov_hami}. However our results show that it is possible to almost entirely re-magnetize a system whose order parameter has been strongly suppressed on a ultra-fast time scale. Notably this re-magnetization occurs even though the double occupation reaches large values, by reversing the dynamics of the latter as well. 


\section{Conclusions}
\label{sec:conclusion}
In the present work we have discussed several forms in which the melting of an initial perfect N\'eel state in a high-dimensional Fermi-Hubbard lattice can be controlled by external periodic driving. Using the recently-introduced non-equilibrium dynamical mean-field theory with a Hamiltonian-based matrix product impurity solver, we have performed the time evolution of driven systems with a computational effort similar to that of quenched Hamiltonians. In addition, insights from Floquet theory have allowed us to understand the underlying mechanisms responsible for the melting of the magnetic order in each scenario considered.

We focused on three different driving regimes. First we observed how high-frequency driving suppresses the hopping and the exchange interaction, slowing down the melting of the magnetic order. For weak Coulomb repulsion this corresponds to switching from quasiparticle-governed magnetic melting to the scrambling of magnetic order due to mobile charge carriers. Second we considered resonant driving, and observed how for strong interactions it leads to a very fast magnetization decay and a large enhancement of double occupations. This corresponds to switching from melting due to charge carriers to quasiparticle-governed dynamics. In addition, we showed how resonant driving can effectively induce zero or attractive interactions on an ultra-fast time scale, where the latter might be relevant for non-equilibrium superconductivity. Third we considered in-gap off-resonant driving, where for large interactions the driving-induced double occupation is compensated by the suppression of the hopping, while the exchange interaction is weakly affected. Thus the underlying magnetic melting mechanism is the same of the non-driven system, governed by a strong spin-charge coupling. It is important to stress that these results illustrate how driving with the same amplitude $A$, in particular those for which $\mathcal{J}_0(A)=0$, can lead to completely different consequences for distinct frequency regimes. In this case hopping processes are suppressed, but while it leads to dynamical localization at high frequency, it maximizes the creation of charge excitations at resonance.

Gathering ideas from the different driving regimes we presented combined schemes of magnetic control, namely the stabilization of states with determined magnetic order parameter or double occupation, and the induction of dynamics reversal in a short time scale. Our work thus indicates that ultra-fast periodic modulation might provide a viable mechanism to control antiferromagnetic order in strongly interacting systems. This could be observed experimentally in condensed matter systems by excitation with THz light pulses~\cite{mankowski2016rep}, or in cold atomic gases~\cite{mazurenko2017nat} in an oscillating underlying optical lattice~\cite{eckardt2017rmp,goldman2016nat}.  

Finally we emphasize that our work manifests the power of the NE-DMFT algorithm to successfully describe strongly-driven many-body systems in different interacting regimes, without resorting to perturbation theory. It also shows that conventional MPS time evolution based on Trotterized two-site unitary operations works as well as the previously-used Krylov-space algorithms~\cite{alexander2014prb,balzer2015prx}, since with both methods similar timescales are reached with comparable computational effort. This motivates the use of our approach for analyzing how to manipulate and enhance, by external periodic driving, different types of ordered states in high-dimensional lattices such as charge-density waves and superconducting phases, for repulsive~\cite{jonathan2016} and attractive~\cite{kitamura2016prb,nocera2017pra} fermionic models. 

\section*{Acknowledgements}
The authors would like to acknowledge the use of the University of Oxford Advanced Research Computing (ARC) facility in carrying out this work. \url{http://dx.doi.org/10.5281/zenodo.22558}. This research is partially funded by the European Research Council under the European Union's Seventh Framework Programme (FP7/2007-2013)/ERC Grant Agreement no. 319286 Q-MAC. This work was also supported by the EPSRC National Quantum Technology Hub in Networked Quantum Information Processing (NQIT) EP/M013243/1, and the EPSRC projects EP/P009565/1 and EP/K038311/1. J.J.M.A. and F.J.G.R. acknowledge financial support from Facultad de Ciencias at UniAndes-2015 project "Quantum control of non-equilibrium hybrid systems-Part II", and F. Rodr\'iguez, L. Quiroga and J. Coulthard for interesting discussions. J.J.M.A. also acknowledges S. Al-Assam and J. Kreula for their help during the development of the DMFT code.

\bibliographystyle{andp2012}
\bibliography{dmft_bib}

\end{document}